\documentclass[letterpaper, oneside, 12pt]{article}
\usepackage{amsmath}
\usepackage{graphicx}%
\usepackage{amsfonts}%
\usepackage{amssymb}
\usepackage{overpic}
\usepackage{subfigure}
\usepackage{rotating}
\usepackage{epstopdf}
\usepackage{url}
\usepackage{color}
\usepackage{setspace}
\usepackage{fullpage}
\usepackage{multirow}
\usepackage{lineno}

\newcommand{\bA}{ {\boldsymbol A} }

\newcommand{\bC}{ {\boldsymbol C} }

\newcommand{\bD}{ {\boldsymbol D} }

\newcommand{\bI}{ {\boldsymbol I} }

\newcommand{\bK}{ {\boldsymbol K} }

\newcommand{\bm}{ {\boldsymbol m} }

\newcommand{\bO}{ {\boldsymbol O} }

\newcommand{\bs}{ {\boldsymbol s} }

\newcommand{\bt}{ {\boldsymbol t} }

\newcommand{\bu}{ {\boldsymbol u} }

\newcommand{\bV}{ {\boldsymbol V} }
\newcommand{\bw}{ {\boldsymbol w} }

\newcommand{\bx}{ {\boldsymbol x} }

\newcommand{\bbeta}{ {\boldsymbol \beta} }

\newcommand{\bGamma}{ {\boldsymbol \Gamma} }

\newcommand{\bphi}{ {\boldsymbol \phi} }

\newcommand{\bmu}{ {\boldsymbol \mu} }

\newcommand{\bet}{ {\boldsymbol \eta} }

\newcommand{\bSigma}{ {\boldsymbol \Sigma} }
\newcommand{\btheta}{ {\boldsymbol \theta} }

\newcommand{\bzero}{ {\boldsymbol 0} }

\newcommand{\given}{\,|\,}

\newcommand{\qed}{\nobreak \ifvmode \relax \else
      \ifdim\lastskip<1.5em \hskip-\lastskip
      \hskip1.5em plus0em minus0.5em \fi \nobreak
      \vrule height0.75em width0.5em depth0.25em\fi}
\begin{document}

\setcounter{page}{0}
\thispagestyle{empty}

\begin{center}

{\Large \textbf{Dynamic spatial regression models for space-varying forest stand tables}}
\vspace{.1in}

\vspace{.1in}

{\large \textsc{Andrew O. Finley, Sudipto Banerjee,\\Aaron R. Weiskittel, Chad Babcock, and Bruce D. Cook}}\footnote[1]{
Andrew O. Finley, Departments of Geography and Forestry, Michigan State University, East Lansing, U.S.A.;
Sudipto Banerjee, Department of Biostatistics, UCLA School of Public Health, Los Angeles, U.S.A.;
Aaron R. Weiskittel, School of Forest Resources, University of Maine, Orono, U.S.A.;
Chad Babcock, School of Environmental and Forest Sciences, University of Washington, Seattle, U.S.A.;
Bruce D. Cook, Biospheric Sciences Laboratory, National Aeronautics and Space Administration, Greenbelt, U.S.A.}

\vspace{.5in} Corresponding author:  Dr. A. O. Finley\\
{\em Department of Forestry, Michigan State University,\\}
\vspace{-.15in} {\em East Lansing, Michigan, 48824, U.S.A.}

\vspace{.5in} telephone: (517) 432-7219

\vspace{-.15in} email: finleya@msu.edu
\vspace{.1in}

\end{center}

\newpage

\begin{center}
{\Large \textbf{Dynamic spatial regression models for space-varying forest stand tables}}
\end{center}

\textsc{abstract:}
Many forest management planning decisions are based on information about the number of trees by species and diameter per unit area. This information is commonly summarized in a \emph{stand table}, where a stand is defined as a group of forest trees of sufficiently uniform species composition, age, condition, or productivity to be considered a homogeneous unit for planning purposes. Typically information used to construct stand tables is gleaned from observed subsets of the forest selected using a probability-based sampling design. Such sampling campaigns are expensive and hence only a small number of sample units are typically observed. This data paucity means that stand tables can only be estimated for relatively large areal units. Contemporary forest management planning and spatially explicit ecosystem models require stand table input at higher spatial resolution than can be affordably provided using traditional approaches. We propose a dynamic multivariate Poisson spatial regression model that accommodates both spatial correlation between observed diameter distributions and also correlation between tree counts across diameter classes within each location. To improve fit and prediction at unobserved locations, diameter specific intensities can be estimated using auxiliary data such as management history or remotely sensed information. The proposed model is used to analyze a diverse forest inventory dataset collected on the United States Forest Service Penobscot Experimental Forest in Bradley, Maine. Results demonstrate that explicitly modeling the residual spatial structure via a multivariate Gaussian process and incorporating information about forest structure from LiDAR covariates improve model fit and can provide high spatial resolution stand table maps with associated estimates of uncertainty.



\section{Introduction}\label{intro}
Sustainable forest management decisions require detailed information about the number and sizes of trees in a forest. Traditionally, this information is summarized in a \emph{stand table} that reports number of trees by some characteristic (e.g., species most commonly, grade, condition) and diameter class per unit area. Stand tables have a long history in forestry because they are an effective way to summarize forest inventory data and inform silvicultural prescriptions (Husch et al. 2003).  Most operational forest inventories use a probability-based sampling design to identify subsets of trees within forest stands to measure. Stand tables are then constructed using observed tree counts per unit area within diameter classes of some convenient increment, e.g., 1 or 2 cm, typically measured at breast height 1.37 m, i.e., diameter at breast height (DBH). Such inventory approaches are expensive and, hence, data used to estimate stand tables are typically spatially and temporally sparse. The traditional design-based estimators used in these settings are unable to generate spatially explicit diameter class distributions with associated uncertainty needed to inform many contemporary management decisions. Spatially explicit stand table estimates are key inputs to terrestrial ecosystem models such as the Ecosystem Demography Model (Medvigy et al., 2009; Medvigy and Moorcroft, 2012) that predicts ecosystem structure (e.g., above and below-ground biomass, vegetation height and forest basal area, and soil carbon stocks) and corresponding ecosystem fluxes (e.g., net primary productivity, net ecosystem production, and evapotranspiration) from climate, soil, and land-use inputs.

There is a long history of using statistical probability density functions to estimate tree diameter distributions using sample data of tree count by DBH class (Weiskittel et al., 2011). The most common distributions used include the Weibull (Schreuder and Swank, 1974), Beta (Maltamo et al., 1995), and Johnson's S$_b$ (Fonseca et al., 2009), while a variety of other distributions like the logit-logistic (Wang and Rennolls, 2005) and Gamma (Hafley and Schreuder, 1977) have been applied to a lesser extent. The parameters for these distributions have been estimated using a variety of approaches including Bayesian (Green et al., 1994), maximum likelihood (Robinson, 2004), and method of moments (Burk and Newberry, 1984), which can strongly influence the accuracy and precision of the derived values (Poudel and Cao, 2013). Along with parametric techniques, a variety of semi- and non-parametric approaches have been tested including finite mixture models (Zhang et al., 2001; Liu et al., 2002), percentile-based (Borders et al., 1987), and $k$-most similar neighbor imputation (Maltamoa et al., 2009). 

Although different parametric and non-parametric approaches can produce similar results (Bollands\r{a}s et al., 2013), both procedures have some important shortcomings. First, most approaches, particularly the parametric ones, predict the relative frequency when the absolute frequency is needed by forest managers. This means that total tree density must either be predicted or measured to scale the relative frequency distribution. Second, extending predictions to new unsampled stands can be problematic because it requires either predicting the parameters of the probability density functions from stand-level attributes or an extensive calibration dataset. Third, most approaches are unable to incorporate covariates to help explain variability in the diameter class distribution and improve prediction. Finally, except for the $k$-most similar neighbor imputation, all the methods cited above ignore the spatial correlation and correlation within observed diameter distributions, which has been shown to be strong in many forest settings (Berhe, 1999; Salas et al., 2010).

The current manuscript seeks to address the shortcomings noted above by developing a framework to jointly model total tree density and diameter distribution, while accommodating spatial dependence across locations and dependence in tree counts across DBH classes. The proposed framework can also incorporate covariates, such as management history or proxies of structural complexity derived from remotely sensed data, and deliver spatially explicit stand table predictions with associated estimates of uncertainty.

The format of the manuscript is as follows. Section~\ref{methods} develops the proposed framework for modeling space-varying diameter class distributions including details on an adaptive Markov chain Monte Carlo (MCMC) algorithm used for parameter estimation and prediction at new locations. Section~\ref{Sec: data_analysis} illustrates our proposed model using a forest inventory dataset. Finally, Section~\ref{Sec: summary} concludes the manuscript with a focus on future work.

\section{Methods}\label{methods}

\subsection{Model}\label{Sec: Model_Estimation}
Let ${\cal D} \subset \Re^2$ be a connected subset of the $2$-dimensional Euclidean space and let $\bs\in {\cal D}$ be a generic point in ${\cal D}$. Our outcome variable at location $\bs$, $y_{ij}(\bs)$, denotes the number of trees belonging to species $i$, where $i=1,2,\ldots,q$, corresponding to diameter class $j$ among a succession of non-overlapping tree diameter classes $\{(0,1], (1,2], \ldots, (m-1, m]\}$. For each location, species and diameter class a $p_{ij}\times 1$ vector $\bx_{ij}(\bs)$ that includes an intercept and $p_{ij}-1$ covariates is also recorded.

We intend to capture three different types of association: ($i$) spatial association exhibited by observations from a location being similar to those from neighboring locations; ($ii$) association exhibited by observations from adjacent diameter classes; and, ($iii$) association among observations arising from different species of trees, perhaps attributable to underlying biological processes. We model ($i$) using a spatial process over an Euclidean space, in particular a Gaussian process specified by a spatial covariance function. For ($ii$), we assume a Markovian (or autoregressive) structure across the diameter classes. This is not dissimilar to the rich class of space-time dynamic models (see, e.g., Gelfand et al., 2005; Finley et al., 2012), except that time is now replaced by diameter class. Finally, for ($iii$), we use an unstructured variance-covariance matrix to model the between-species association. Furthermore, we assume that the impact of the covariates on the intensity is specific to the diameter class and the species of tree corresponding to that observation. Thus, the relationship between the covariates and the number of trees depends upon the spatial location, the species of the tree and the diameter class. We propose the following hierarchical spatial Poisson regression model for each location $\bs$, species $i=1,2,\ldots,q$ and diameter class $j=1,2,\ldots,m$, 
\begin{align}\label{Eq: Dynamic_Diameter_Class_Model}
y_{ij}(\bs) &\stackrel{ind}{\sim} Poi(\lambda_{ij}(\bs))\; ;\quad \log\lambda_{ij}(\bs) = \bx_{ij}(\bs)'\bbeta_{ij} + w_{ij}(\bs)\; ;\nonumber \\
\bbeta_{ij} & = \bbeta_{i,j-1} + \bet_{ij},\quad \bet_{ij}\stackrel{ind}\sim N(\bzero,\bSigma_{\eta,j})\; ;\nonumber \\
w_{ij}(\bs) &= w_{i,j-1}(\bs) + u_{ij}(\bs),\quad \bu_{j}(\bs) \stackrel{ind}{\sim} MVGP\left(\bzero, \bC_{j}(\cdot,\btheta_{j})\right)\;.
\end{align} 
Here, $\log\lambda_{ij}(\bs) = \bx_{ij}(\bs)'\bbeta_{ij} + w_{ij}(\bs)$, $\bbeta_{ij}$ is a $p_{ij}\times 1$ vector of regression coefficients, $\bu_j(\bs) = (u_{1j}(\bs), u_{2j}(\bs), \ldots, u_{qj}(\bs))'$ is a $q\times 1$ vector following a multivariate Gaussian process, $MVGP(\bzero, \bC_{j}(\cdot,\btheta_{j}))$, with cross-covariance function $\bC_{j}(\cdot;\btheta_{j})$. A comprehensive treatment of multivariate Gaussian processes and cross-covariance functions can be found in Gelfand and Banerjee (2010) or in Banerjee et al. (2014). 

Motivated by the so-called ``linear model of coregionalization'' (Wackernagel, 2010), we specify $\bC_j(\bs,\bt;\btheta_j) = \bA_j\bD(\bphi_j)\bA_j'$, where $\btheta_j = \{\bA_j,\bphi_j\}$, $\bA_j$ is an unknown $q\times q$ lower-triangular matrix and $\bD(\bphi_j)$ is diagonal. The univariate spatial correlation function $\rho_{ij}(\bs,\bt;\phi_{ij})$ is the $i$-th diagonal entry in $\bD(\bphi_j)$, where $\bphi_j = \{\phi_{1j},\phi_{2j},\ldots,\phi_{qj}\}$. An exponential correlation function is often used to define the spatial correlation structure, e.g., $\rho_{ij}(\bs,\bt;\phi_{ij})=\exp(-\phi_{ij}\|\bs-\bt\|)$, where $\|\bs-\bt\|$ is the Euclidean distance between the sites $\bs$ and $\bt$. Of course, any other \emph{valid} spatial correlation function, such as the Mat\'ern could be used (see, e.g., Banerjee et al., 2014). The elements of $\bA_j\bA_j'$ provide the variances and covariances among the $q$ species, specific to each diameter class. 

We further assume $\bbeta_{i0} \stackrel{iid}{\sim}N(\bm_0, \bSigma_0)$ and $w_{i0}(\bs) \equiv 0$ for every $i=1,2,\ldots,q$, where $\bm_0 = \bzero$ and $\bSigma_0$ is a diagonal matrix with very large diagonal elements. Customarily, a flat prior is assigned to each $\bbeta_{i0}$. The prior specifications for a Bayesian hierarchical model are completed by assigning hyper-priors to each $\bSigma_{\eta,j}$ and $\btheta_j$. The former captures possible association among the regression coefficients, specific to each diameter class, and is typically assigned an inverse-Wishart (IW) prior independent across diameter classes. For $\btheta_j = \{\bA_j,\bphi_j\}$, one assigns independent priors for $\bA_j$ and each element of $\bphi_j$. The prior for $\bA_j$ can be induced from assigning an IW prior on $\bA_j\bA_j'$ or by modeling, independently, the diagonal elements as log-normal and the remaining elements as normal. The prior for the elements of $\bphi_j$, i.e., each $\phi_{ij}$, is usually taken to be uniform distributions whose bounds are obtained by taking into account spatial domain considerations (e.g., the maximum inter-site distance). The precise specifications used are discussed in a subsequent section. 

The model in (\ref{Eq: Dynamic_Diameter_Class_Model}) is envisioned at any \emph{arbitrary} location $\bs\in {\cal D}$. It is a well-defined \emph{process model} in that it yields a legitimate joint probability model for any finite collection of spatial locations in ${\cal D}$. To be precise, let ${\cal S} = \{\bs_1,\bs_2,\ldots,\bs_n\}$ be a set of $n$ locations in ${\cal D}$, where the outcome and covariates have been recorded. In practice, the estimation of model parameters is usually very robust to the above hyper-prior specifications. Using notations similar to Gelman et al. (2013), we obtain the following joint distribution for the parameters and the data,
\begin{align}\label{Eq: Hier_Multivariate_Poisson}
& \prod_{j=1}^m p(\btheta_j) \times \prod_{j=1}^m IW(\bSigma_{\eta,j}\given r_{\eta}, \Upsilon_{\eta}) \times \prod_{i=1}^q N(\bbeta_{i0}\given \bm_0,\bSigma_0)\times \prod_{i=1}^q\prod_{j=1}^m N(\bbeta_{ij} \given \bbeta_{i,j-1},\bSigma_{\eta,j}) \nonumber \\
& \qquad \times  \prod_{j=1}^m N(\bw_j \given \bw_{j-1}, \bSigma_j(\btheta_j)) \times \prod_{i=1}^q\prod_{j=1}^m\prod_{k=1}^{n} \lambda_{ij}(\bs_k)^{y_{ij}(\bs_{k})}e^{-\lambda_{ij}(\bs_k)}\; ,
\end{align} 
where $\bw_j = (\bw_j(\bs_1)',\bw_j(\bs_2)',\ldots,\bw_j(\bs_n)')'$ is an $nq\times 1$ vector of spatial random effects, each $\bw_j(\bs_k)$ is $q\times 1$ with elements $w_{ij}(\bs_k)$, $\bSigma_j(\btheta_j)$ is an $nq\times nq$ spatial covariance matrix whose $(k,l)$-th block is the $q\times q$ cross-covariance matrix $\bC_j(\bs_k,\bs_l;\btheta_j)$. Some special cases are immediate. If the Markovian dependence on the $\bbeta_{ij}$'s is undesirable, then we can simply replace the term $\prod_{i=1}^qN(\bbeta_{i0}\given \bm_0,\bSigma_0)\times \prod_{i=1}^q\prod_{j=1}^m N(\bbeta_{ij} \given \bbeta_{i,j-1},\bSigma_{\eta,j})$ in (\ref{Eq: Hier_Multivariate_Poisson}) by $\prod_{i=1}^q\prod_{j=1}^m N(\bbeta_{ij} \given \bmu_j,\bSigma_{\eta,j})$ and each $\bmu_j$ is customarily set to $\bzero$. Also, a flat prior on $\bbeta_{ij}$'s would simply set $\bSigma_{\eta,j}^{-1} = \bO$ (the null matrix) for each $j$. This is equivalent to removing all terms involving $\bSigma_{\eta,j}$ and the $\bbeta_{ij}$'s (including $\bbeta_{i0}$) from (\ref{Eq: Hier_Multivariate_Poisson}), except those appearing in the $\lambda_{ij}(\bs_k)$'s.

\subsubsection{Prediction}
Spatial prediction proceeds in a posterior predictive fashion using posterior samples of $\bbeta_{ij}$'s, $\btheta_j$'s and $\bw_j$'s in (\ref{Eq: Hier_Multivariate_Poisson}). Let $y_{ij}(\bs_0)$ be the random variable denoting the unknown value of the outcome at an arbitrary location $\bs_0$ for any species type $i$ and diameter class $j$. We draw, one-for-one, the $q\times 1$ random effect vector $\bw_{j}(\bs_0)$ from a normal distribution with mean vector and variance-covariance matrix 
\[
\sum_{l=1}^j\bK_l(\bs_0;\btheta_l)'\bSigma_l(\btheta_l)^{-1}\left(\bw_l - \bw_{l-1}\right)\, \mbox{ and }\, \sum_{l=1}^j\left\{\bC_l(\bs_0,\bs_0) - \bK_l(\bs_0;\btheta_l)'\bSigma_l(\btheta_l)^{-1}\bK_l(\bs_0;\btheta_l)\right\}.
\]
Here, $\bK_l(\bs_0;\btheta_l)$ is $nq\times q$ with $i$-th block given by the $q\times q$ matrix $\bC_l(\bs_0,\bs_i;\btheta_l)$. Then, for a known $\bx_{ij}(\bs_0)$, we plug in the samples of $w_{ij}(\bs_0)$ and the posterior samples of $\bbeta_{ij}$ to obtain samples of $\log\lambda_{ij}(\bs_0) = \bx_{ij}(\bs_0)'\bbeta_{ij} + w_{ij}(\bs_0)$ and, subsequently, of $y_{ij}(\bs_0) \sim Poi(\lambda_{ij}(\bs_0))$. The resulting samples of $y_{ij}(\bs_0)$ constitute the posterior predictive distribution of $y_{ij}(\bs_0)$.

\subsection{Implementation}\label{Sec: Estimation}
The joint posterior distribution for the model parameters is proportional to (\ref{Eq: Hier_Multivariate_Poisson}) but intractable otherwise, so we sample from the posterior distribution using MCMC algorithms (see, e.g., Robert and Casella, 2004). Such algorithms can be built for spatial dynamic linear models by extending ideas laid out in numerous earlier papers including, but not limited to, De Jong (1989), Koopman (1993), Shepard and Pitt (1997), and Gamerman (1998). Our first stage specification, i.e., the likelihood, is Poisson, which precludes analytically tractable expressions from integrating out $\bw$ in (\ref{Eq: Hier_Multivariate_Poisson}). Therefore, our sampler needs to operate on a high-dimensional space. A Gibbs-sampler with random-walk Metropolis updates that draws from the full-conditional distribution of each parameter (vector) is easy to program. However, the Metropolis steps are awkward to tune and typically suffer from high autocorrelations resulting in very slow convergence.  
For the analysis presented in Section~\ref{Sec: data_analysis}, we employ an adaptive Metropolis-within-Gibbs algorithm outlined in Roberts and Rosenthal (2009) to draw posterior samples from (\ref{Eq: Hier_Multivariate_Poisson}). This adaptively (and automatically) tunes the Metropolis steps and offers substantially superior performance in high-dimensional hierarchical models such as (\ref{Eq: Hier_Multivariate_Poisson}).  

As in the usual Metropolis sampling algorithms, we propose updating the $l$-th parameter, possibly a $k\times 1$ vector, by adding a $N_k(\bzero,\bGamma_l)$ increment to the current value, which is then accepted or rejected according to the usual Metropolis ratio. The $k\times k$ matrix $\bGamma_l$ tunes the proposal and is assumed diagonal with entries $\gamma_{lj}^2$, $j=1,2,\ldots,k$. However, unlike the usual Metropolis sampler, we do not leave the tuning parameter $\bGamma_l$ fixed. Instead, we begin with $\bGamma_l=\bI_k$ and then update each $\bGamma_l$ after every $b$-th batch of $50$ iterations. For each $j=1,2,\ldots,k$, we change $\bGamma_l$ according to $\gamma_{lj} = \exp(\log(\gamma_{lj}) + \delta(b))$ if the acceptance rate in the $b$-th batch for the $l$-th parameter exceeds $0.44$, or to $\gamma_{lj} = \exp(\log(\gamma_{lj}) - \delta(b))$ if was less than $0.44$, where $\delta(b)=\min\{0.01,1/\sqrt{b}\}$. Since $\delta(b)\to 0$ as $b\to\infty$, the adaptive MCMC satisfies ergodicity ensuring much faster convergence (Roberts and Rosenthal, 2009). For scalar parameters, we simply set $k=1$. In theory, this algorithm works for all target densities that are log-concave outside of a bounded interval. While the target densities arising from (\ref{Eq: Hier_Multivariate_Poisson}) are not, strictly speaking, log-concave in the process parameters, i.e., the $\btheta_j$'s, this algorithm seems to perform very effectively in practice. 


More specifically, we use the above adaptive Metropolis steps to update each parameter in (\ref{Eq: Hier_Multivariate_Poisson}) from its full conditional distribution. We use element-wise scalar updates for the entries in the $\bbeta_{i,0}$'s, $\bbeta_j$'s and $\bphi_j$'s. We also use scalar updates for each lower-triangular entry in $\bA_j$ and in the Cholesky square root of $\bSigma_{\eta,j}=\bV_j\bV_j'$. To avoid autocorrelations, the spatial effects for each diameter class, i.e. the $\bw_j$'s are updated as $nq\times 1$ vectors. The diagonal entries in $\bV_j$ and $\bA_j$, which are positive to ensure identifiability, are log transformed for the proposal. The elements in the $\bphi_j$'s are also positive and modeled with a uniform prior and are conveniently mapped to the whole real line using a logit transformation. Necessary Jacobian adjustments are included in (\ref{Eq: Hier_Multivariate_Poisson}).

\subsection{Model selection and prediction}\label{Sec: Prediction}
Several sub-models of (\ref{Eq: Dynamic_Diameter_Class_Model}) are considered for the forest data analysis described in Section~\ref{Sec: data_analysis}. Here, we use a model fit criterion and out of sample prediction performance to rank the candidate models. Model fit is assessed using the \emph{Deviance Information Criterion} (DIC) (Spiegelhalter et al., 2002). This criterion is the sum of the Bayesian deviance (a measure of model fit) and the (effective) number of parameters $p_D$ (a penalty for model complexity). Here, lower DIC indicates \emph{better} fit. Out of sample observed versus predicted values within each diameter class are compared using proper and strictly proper scoring rules. Here, we consider the logarithmic (LogS), squared error (SES), and Dawid-Sebastiani (DSS) scoring rules defined in Equations 5, 9, 11, respectively, in Czado et al. (2009). For all three rules, lower values indicate improved predictive performance.

\section{Illustration}\label{Sec: data_analysis}

\subsection{Study area and data}
The Penobscot Experimental Forest (PEF) is a 1600 ha tract of Acadian mixed-species forest located in Bradley, Maine (44$^\circ$ 52' N, 68$^\circ$ 38' W) (Figure \ref{map}). The average annual temperature and precipitation near Bradley is 6$^\circ$C and 110 cm respectively. Species composition on the PEF is a mixture of coniferous and deciduous trees including spruce (\textit{Picea spp.}), balsam fir (\textit{Abies balsamea}), red maple (\textit{Acer rubrum}), birch (\textit{Betula spp.}), and aspen (\textit{Populus spp.}), among others. The PEF is dominated by conifer species, the majority of which are shade-tolerant. Since the 1950s, routine management and monitoring has occurred in the PEF's 50$+$ management units (MUs). With over 600 $\sim$0.2 ha georeferenced permanent sample plots (PSPs) currently established and scheduled for remeasurement on an approximate ten-year cycle (and immediately pre- and post-harvest), the PEF has a wealth of data available for assessing forest composition and structural characteristics such as diameter class distributions (Brissette et al., 2012).  

\subsubsection{Management on the PEF}
Different silvicultural treatments are implemented within each MU, e.g., unregulated harvest, shelterwood, diameter limit cutting, natural regeneration. Some specific MUs are highlighted in Figure~\ref{map} and referenced in Section~\ref{results} for interpreting model results. MU 8 was commercially clearcut in 1983. MU 4 received a diameter-limit harvest in 1994. Harvest DBH thresholds were 14.0 cm for balsam fir, 24.1 cm for spruce and hemlock (\textit{Tsuga canadensis}), 26.7 cm for white pine, and 19.1 cm for cedar (\textit{Thuja occidentalis}). MU 9 and 16 are managed under a five-year cutting cycle where the structural goal is to retain a cross-sectional area of 14.1 m$^2$/ha of trees greater than 11.4 cm DBH. MU 23 is undergoing a three-stage shelterwood harvest and was last entered in 2007. MU 32 is a mature, natural area that serves as a reference stand for the PEF with limited harvesting or other management actions conducted since 1954. Additional information about the silvicultural practices for these and other PEF stands are given by Sendak et al. (2003) and Hayashi et al. (2014).

\subsubsection{Observed diameter class distributions}\label{PSP}
For this analysis, we considered the 430 PSPs sampled after 2005. This date cutoff was used to minimize error due to temporal misalignment between remotely sensed covariates described in Section~\ref{gliht} and field measurement data. Trees with DBH of 12.7 cm and greater were measured on all PSPs. For each PSP, tree count by 2.54 cm diameter increments were summarized for both shade-tolerant and shade-intolerant conifer species with 55.88 set to the maximum DBH. Half of the PSPs were used to fit the candidate models, red points in Figure~\ref{map}, and the other half were reserved as holdout data to assess prediction performance. Thus, $q=2$, $m$=18, and $n$=215 in (\ref{Eq: Dynamic_Diameter_Class_Model}) and (\ref{Eq: Hier_Multivariate_Poisson}). These summaries were then scaled to reflect per ha tree counts. Solid and open circles in Figure~\ref{fitted} show observed tree count for the two species groups for six example PSPs. The shape of these distributions is indicative of the different harvesting precipitations applied to the MUs. For example, Figure~\ref{h1} shows a typical diameter distribution for a PSP within MU 32 which is an unmanaged mature forest stand---characterized by a structurally diverse distribution with many large diameter trees. In contrast Figure~\ref{h2} exemplifies a diameter distribution associated with a regenerating clearcut. Here, with the stand clearing harvest in 1983, this MU shows a clear peak of even-aged shade-tolerant trees of about 17.78 cm and no trees larger than $\sim$33 cm. The distributions in Figure~\ref{fitted} also show the paucity of shade-intolerant conifers versus the abundance of shade-tolerant species across the PEF. Specifically, there were 7921 shade-tolerant versus 853 shade-intolerant trees on the 215 observed PSPs. 

\subsubsection{LiDAR acquisition and preparation}\label{gliht}
Data from the National Aeronautics and Space Administration (NASA) LiDAR, Hyperspectral, and Thermal (G-LiHT; Cook et al., 2013) sensor were collected over the extent of the PEF in 2012. G-LiHT is a portable airborne system developed by NASA Goddard Space Flight Center that simultaneously maps the composition, structure, and function of terrestrial ecosystems. The G-LiHT airborne laser scanner (VQ-480, Riegl Laser Measurement Systems, Horn, Austria) uses a 1550 nm laser that provides an effective measurement rate of up to 150 kHz along a 60$^\circ$ swath perpendicular to the flight direction. At a nominal flying altitude of 335 m, each laser pulse has a footprint approximately 10 cm in diameter and is capable of producing up to 8 returns. Pseudo-waveforms were created for the PEF by aggregating G-LiHT LiDAR returns and weighting return heights using a Gaussian shaped 25 m diameter footprint (Blair and Hofton, 1999). This processing resulted in 29571 pseudo-waveforms covering the extent of the PEF. Percentile height variables at 5\% intervals between 5\% and 100\% where calculated for each pseudo-waveform (Figure~\ref{lidar}). These variables represent the canopy height below which the given percent of laser energy was returned and are useful for describing the vertical structure of forest biomass at a given location (see, e.g., Gobakken and N{\ae}sset, 2005; N{\ae}sset and Gobakken, 2005). For the subsequent regression analysis we chose the 5, 25, 50, and 95-th percentile heights for use as covariates. These percentiles were chosen because they were not highly correlated and provide information about the lower, mid, and upper canopy forest structure.

\subsection{Candidate models and computing}\label{candidates}
Stem count per ha by diameter class for shade-tolerant and shade-intolerant species was modeled using three different specifications of (\ref{Eq: Dynamic_Diameter_Class_Model}). The candidate models were: ($i$) \emph{non-spatial with LiDAR}, which includes the LiDAR covariates but sets the multivariate spatial random effects $\bw_j$'s to zero; $ii$) \emph{spatial without LiDAR}, which includes the multivariate spatial random effects but does not include the LiDAR covariates; and, $iii$) \emph{spatial with LiDAR}, which includes the LiDAR covariates and the multivariate spatial random effects. For all candidate models, we made the simplifying assumption that the $\bSigma_{\eta,j}$'s were common across diameter classes. This is a reasonable assumption because we do not expect the relationship between vertical vegetation structure and LiDAR returns to vary by diameter class or species.

The specification for (\ref{Eq: Hier_Multivariate_Poisson}) is completed by assigning hyper-priors to each parameter's prior distribution. We assigned $\bm_0=\bzero$ and $\bSigma_0=1000\bI_p$ for $\bbeta_{i0}$'s normal, where $\bI_p$ is $p\times p$ identity matrix and $p$ is the number of model covariates including the intercept. The regression coefficients' $p\times p$ variance-covariance matrix $\bSigma_{\eta}$ was assigned an IW with degrees of freedom $r_{\eta}=p+1$ and scale matrix $\Upsilon_{\eta}=0.01\bI_p$. For the two spatial models, we used an exponential spatial correlation for the $\rho_{ij}(\cdot)$'s, with spatial decay parameters following a uniform distribution with support between 0.1--6 km (where 6 km is approximately the maximum inter-site distance between PSPs). Each $q\times q$ spatial variance-covariance matrix $\bGamma_j=\bA_j\bA_j'$ was assigned an IW with $r_{\Gamma}=q+1$ and scale matrix $\Upsilon_{\Gamma}=0.01\bI_q$.

The MCMC sampler described in Section~\ref{Sec: Estimation} was implemented in C++ and used Intel’s Math Kernel Library threaded BLAS and LAPACK routines for efficient matrix operations. The sampler was run on a Linux workstation using an Intel Xeon 10 core processor with hyper-threading. Posterior inference was based on three MCMC chains run for 75,000 iterations each with the first 15,000 iterations discarded as burn-in. {Each chain was given different starting values and chain mixing and convergence was assessed using trace plots and the Gelman-Rubin diagnostic (Gelman and Rubin, 1992). The spatial model with covariates leveraged $\sim$10 cores simultaneously for matrix operations and required $\sim$15 hours to run a single chain. A Cholesky decomposition of each DBH class's $nq\times nq$ spatial covariance matrix, needed for evaluating the likelihood in (\ref{Eq: Hier_Multivariate_Poisson}), was the most computationally demanding step in the sampler.

\subsection{Results and discussion}\label{results}
Table~\ref{dic} provides the DIC and associated metrics used to rank the candidate models. Despite the larger effective number of parameters penalty of 214, the lower value of DIC for the spatial model with covariates suggests that inclusion of LiDAR information and addition of the dynamic spatial random effects improves model fit over that achieved by the covariates or spatial random effects alone. Scoring rule results for the out of sample prediction are given in Figure~\ref{scores}.  Here, again, lower values indicated improved prediction. The scoring rules generally agree, but do show some differences. Generally, the spatial with LiDAR model performs marginally better, and more consistent across the diameter classes, than the other two candidate models. Subsequent results and discussion focus on the full spatial model with LiDAR covariates. 

Posterior summaries for model parameters are given in the on-line supplementary document. Here, we highlight a few points that support the choice of the spatial with LiDAR covariates model.  Estimates for the diagonal elements of the $m$ $\bGamma_j$'s ranged from 0.03 to 0.25 for shade-tolerant and 0.01 to 0.27 for shade-intolerant species. The inter-species cross-correlations, i.e., off-diagonal elements of the $\bGamma_j$'s converted to correlations, ranged from -0.58 to 0.91 with about half of these correlations possessing 95\% credible intervals (CI) that exclude zero. These strong correlations suggest the use of a multivariate Gaussian process is warranted and tree count information for one species group might help inform counts for the other group. Estimates of effective spatial ranges (defined here as the distance at which the spatial correlation drops to 0.05) across the diameter classes ranged from 0.1 to 3 km for both species groups, with the longest effective ranges seen in the shade-tolerant $\sim$25.4 to 40.64 cm diameter classes. 

Posterior summaries for the dynamic regression coefficients are provided in Figure~\ref{beta}. The intercepts', i.e., $\beta_0$, inverse relationship with DBH Figure~\ref{b0}, seen in both shade-tolerant and shade-intolerant species, reflects the decrease in stem abundance with increasing DBH---a trend common in all MUs. The regression coefficients associated the LiDAR percentile heights are given in Figures~\ref{b4}-\subref{b1}. The larger width of the 95\% CI for the shade-intolerant versus that of the shade-tolerant species is likely due to the sparseness of those species across the PEF. The exclusion of zero from much of the CI band for all regression coefficients and for both species groups suggests the LiDAR covariates are useful for explaining variability in stem count. For example, the 95th percentile is essentially a measure of stand height. We expect a positive relationship between stand height and number of large diameter trees, i.e., larger DBH trees are generally taller when grown in densely populated stands. This relationship is reflected by the positive values of $\beta_{P95}$, and to some extent $\beta_{P50}$, for larger DBH in Figures~\ref{b4} and \subref{b3}. Conversely, the negative values of $\beta_{P95}$ and $\beta_{P50}$ for smaller diameter trees suggests lower canopy heights are associated with a greater number of small diameter trees which is certainly true in regenerating stands and under some of the silvicultural treatments on the PEF. Similar patterns between vertical stand structure and stem abundance by DBH class are reflected in Figures~\ref{b2} and \subref{b1} although multiple scenarios lead to such relationships.

Observed versus model fitted values with associated CIs for six PSPs are given in Figure~\ref{fitted}. These figures, and those for all other PSPs not shown, suggest the proposed model is able to capture the shape and magnitude of the observed absolute frequency of trees per ha across the diameter class distribution. The multiple local maxima observed in, e.g., Figures~\ref{h1}, \subref{h4}, and \subref{h5}, highlight the often complex forest structure within the MUs and underscores the need to move beyond the common approach of fitting simple parametric distributions.

Following from Section~\ref{Sec: Prediction}, we generated posterior predictive samples from diameter class distributions at each of the 29571 locations where LiDAR was collected across the PEF. Each panel in Figures~\ref{pred1} and \ref{pred2} is a species group specific map of trees per ha posterior predictive distribution median for the given diameter class. Overall these maps reflect the global trend of decreasing tree count with increasing DBH and the prevalence of the shade-tolerant versus shade-intolerant species. The maps also reveal several MU scale characteristics. For example, the 12.7-17.8 cm panels in Figure~\ref{pred1} show a high tree per ha count for MU 23 (reference MU index in Figure~\ref{map}) relative to the other MUs. This flush of regeneration in MU 23 is the result of a multi-stage shelterwood treatment initiated in 1983 and completed in 2007. A similar unimodal distribution indicative of even-aged stand regeneration can be seen in MU 8 which underwent a clearcut in 1983. MUs 9 and 16 show an abundance of shade-tolerant trees in the $\sim$33-45.7 cm diameter classes relative to the other MUs, which is the result of silvicultural prescriptions that favor stocking in larger diameter trees (Hayashi et al., 2014). The control MU 32 has not undergone harvesting in the past $\sim$60 years and, as a result, has a complex diameter distribution with trees occupying nearly all DBH classes and species groups, as seen in Figures~\ref{pred1} and \ref{pred2}. Predictive uncertainty for shade-tolerant and shade-intolerant species are mapped in Figures~\ref{pred1Rng} and \ref{pred2Rng}, respectively. As expected, prediction variability is largest for locations with higher predicted trees per ha, see, e.g., MUs 23, 9, and 16.

The PEF dataset is fairly unique in its level of detail about MU boundaries and recorded history of management activities. We could have certainly added a MU effect or harvesting information to the regression portion of (\ref{Eq: Dynamic_Diameter_Class_Model}) which would likely have improved fit and prediction. For example, we would likely see less smoothing across MU boundaries in the prediction maps. However, since many production forests and those not under management do not typically have such MU level information we opted to use only readily available remotely sensed data to help apportion tree count among the diameter classes. 

The time required to obtain parameter estimates for the full model is a clear hurdle to implementation using several species groups observed over thousands of locations. Such settings are common when we consider state or national scale forest inventory datasets, e.g., the United States Forest Service Forest Inventory and Analysis (FIA) database which contains over 100000 PSPs. Here, again, the computational bottleneck is the iterative decomposition of the $nq\times nq$ spatial covariance matrices. Broadly speaking, modeling large spatial datasets proceed by either exploiting ``low-rank'' models or using sparsity (see, e.g., Datta et al., 2014 for a review of pertinent literature). Several such approaches could be used to approximate the $\bu_j(\bs)$'s in (\ref{Eq: Dynamic_Diameter_Class_Model}) and hence reduce the runtime required for large datasets. A similar issue arises when $q$ is large, e.g., $q$ greater than $\sim$10 and $n$ of even moderate size. In such cases the computational demand for estimating the $m$ $\bA_j$'s as well as the spatial covariance matrix may become prohibitively expensive and one might consider dimension reduction via a spatial factor analysis, i.e., specify the $\bA_j$'s with fewer columns than rows (see, e.g., Lopes and West, 2004; Ren and Banerjee, 2013).

\section{Concluding remarks}\label{Sec: summary}
Application of the proposed dynamic model is novel for predicting space-varying forest stand tables and addresses several shortcomings of previous modeling approaches. The proposed framework accommodates the major sources of diameter class distribution dependencies that arise from underlying biological processes and management history, including: ($i$) spatial association among proximate distributions of tree count; ($ii$) association in tree count between adjacent diameter classes; and, ($iii$) association among species specific tree count within and between diameter distributions. The model was tested on a forest with extensive PSPs and an array of species and stand structures. Results suggested the framework was able to captured the information available and leverage it to produced logical predictions. Model fitted and predicted distributions captured the complex spatial patterns in tree size and species distributions, which most previous approaches ignore. This is important because there is currently a need to produce stand tables at a high spatial and temporal resolution for a variety of purposes (see, e.g., Drury and Herynk, 2011). A critical test of the proposed modeling approach would be extending it to larger geographic domains, which, following from the discussion in Section~\ref{results}, will require tackling issues of dimensionality. 

\section*{References}
\begin{description}
\item Banerjee S, Carlin BP, Gelfand, AE. 2014. \emph{Hierarchical Modeling and Analysis for Spatial Data} Second Edition. Boca Raton, FL: Chapman and Hall/CRC Press.

\item Banerjee S, Finley AO, Waldmann P, Ericsson T. 2010 Hierarchical spatial process models for multiple traits in large genetic trials. \emph{Journal of the American Statistical Association} \textbf{105}:506--521.

\item Berhe L. 1999. Spatial continuity in tree diameter distribution. Swedish University of Agricultural Sciences, Deptarmet of Forest Resource Management Ume\r{a}, Sweden. Working Paper No. 64 1999. 

\item Blair JB, Hofton MA. 1999. Modeling laser altimeter return waveforms over complex vegetation using high-resolution elevation data, \emph{Geophysical Research Letters}, \textbf{26}:2509--2512.

\item Bollands\r{a}s OM, Maltamo M, Gobakken T,  N{\ae}sset E. 2013. Comparing parametric and non-parametric modelling of diameter distributions on independent data using airborne laser scanning in a boreal conifer forest. \emph{Forestry} \textbf{0}:1--9.

\item Borders BE, Souter RA, Bailey RL, Ware KD. 1987. Percentile-based disributions characterize forest stand tables. \emph{Forest Science} \textbf{33}:570--576.

\item Brissette JC, Kenefic LS, Russel MB, Puhlick JJ. 2012. Overstory tree and regeneration data from the ``Silvicultural Effects on Composition, Structure, and Growth'' study at Penobscot Experimental Forest. Newtown Square, PA: USDA Forest Service, Northern Research Station. \url{http://dx.doi.org/10.2737/RDS-2012-0008}. 

\item Burk TE, Newberry JD. 1984. A simple algorithm for moment-based recovery of Weibull distribution parameters. \emph{Forest Science} \textbf{30}:329--332.

\item Czado C, Gneiting T, Held L. 2009. Predictive Model Assessment for Count Data. \emph{Biometrics}, \textbf{65}:1259--1261.

\item Cook BD, Corp LW, Nelson RF, Middleton EM, Morton DC, McCorkel JT, Masek JG, Ranson KJ, Ly V, and Montesano PM. 2013. NASA Goddard's Lidar, Hyperspectral and Thermal (G-LiHT) airborne imager. \emph{Remote Sensing} \textbf{5}:4045--4066.

\item Czado C, Gneiting T, Held L. 2009. Predictive model assessment for count data. \emph{Biometrics} \textbf{65}:1254--1261.

\item Datta A, Banerjee S, Finley AO, Gelfand AE. 2014. Hierarchical nearest-neighbor Gaussian process models for large geostatistical datasets. Under review, preprint \url{http://arxiv.org/pdf/1406.7343v1.pdf}.

\item De Jong P. 1989. Smoothing and interpolation with the state-space model. \emph{Journal of the American Statistical Association} \textbf{84}:1085--1088.

\item Drury SA, Herynk JM. 2011. The national tree-list layer. Gen. Tech. Rep. RMRS-GTR-254. Fort Collins, CO: U.S. Department of Agriculture, Forest Service, Rocky Mountain Research Station. 26 p.

\item Finley AO, Banerjee S, Gelfand, AE. 2012. Bayesian dynamic modeling for large space-time datasets using Gaussian predictive processes. \emph{Journal of Geographical Information Systems} \textbf{14}:29--47. 


\item Fonseca TF, Marques CP, Pachecho C, Parresol BR. 2009. Describing maritime pine diameter distributions with Johnson's SB distribution using a new all-parameter recovery approach. \emph{Forest Science} \textbf{55}:367--373.

\item Gamerman D. 1998. Markov chain Monte Carlo for dynamic generalised linear models. \emph{Biometrika} \textbf{85}:215--227.

\item Gelfand AE, Banerjee S, Gamerman, D. 2005. Spatial process modelling for univariate and multivariate Dynamic Spatial Data. \emph{Environmetrics} \textbf{16}:465--479.

\item Gelfand AE, Banerjee S. 2010. Multivariate spatial process models. In Handbook of Spatial Statistics, eds. P Diggle, M Fuentes, AE Gelfand, and P Guttorp, Boca Raton, FL: Taylor and Francis. 

\item Gelman A, Carlin JB, Stern HS, Rubin DB. 2013. \emph{Bayesian Data Analysis}, 3nd edition. Boca Raton, FL: Chapman and Hall/CRC Press.

\item Gelman A, Rubin, DB. 1992. Inference from iterative simulation using multiple sequences. \emph{Statistical Science} \textbf{7}457--511.

\item Gobakken T. N{\ae}sset E. 2004. Estimation of diameter and basal area distributions in coniferous forest by means of airborne laser scanner data. \emph{Scandinavian Journal of Forest Research} \textbf{19}:529--542.

\item Green EJ, Roesch FA Jr, Smith AFM, Strawderman WE. 1994. Bayesian estimation for the three-parameter Weibull distribution with tree diameter data. \emph{Biometrics} \textbf{50}:254--269. 

\item Hafley WL, Schreuder HT. 1977. Statistical distributions for fitting diameter and height data in even-aged stands. \emph{Canadian Journal of Forest Research} \textbf{7}:481--487.

\item Hayashi, R, Weiskittel A, Sader S. 2014. Assessing the Feasibility of Low-Density LiDAR for Stand Inventory Attribute Predictions in Complex and Managed Forests of Northern Maine, USA. \emph{Forests} \textbf{5}:363--383.

\item Husch B, Beers TW, Kershaw JA. 2003. Forest mensuration, 4th edition. John Willey \& Sons, Hoboken, NJ.

\item Koopman SJ. 1993. Disturbance smoother for state space models. \emph{Biometrika} \textbf{80}:117--126.

\item Lopes HF, West, M. 2004. Bayesian model assessment in factor analysis. \emph{Statistica Sinica} \textbf{14},41--67.

\item Liu C, Zhang L, Davis CJ, Solomon DS, Gove JH. 2002. A finite mixture model for characterizing the diameter distributions of mixed-species forest stands. \emph{Forest Science} \textbf{48}:653--661.

\item Maltamo M, Puumalainen J, P\"aivinen R. 1995. Comparison of Beta and Weibull functions for modelling basal area diameter distribution in stands of Pinus sylvestris and Picea abies. \emph{Scandinavian Journal of Forest Research} \textbf{10}:284--295.

\item Maltamoa M, N{\ae}sseta E, Bollands\r{a}sa OM, Gobakkena T, Packal\'enb P. 2009. Non-parametric prediction of diameter distributions using airborne laser scanner data. \emph{Scandinavian Journal of Forest Research} \textbf{24}:541--553.

\item Medvigy D, Wofsy SC, Munger JW, Hollinger DY, Moorcroft PR 2009. Mechanistic scaling of ecosystem function and dynamics in space and time: Ecosystem Demography model version 2. \emph{Journal of Geophysical Research} \textbf{114}:G01002. 

\item Medvigy D,  Moorcroft PR. 2012. Predicting ecosystem dynamics at regional scales: an evaluation of a terrestrial biosphere model for the forests of northeastern North America. \emph{Philosophical Transactions of the Royal Society B} \textbf{367}222--235. 

\item N{\ae}sset E, Gobakken T. 2005. Estimating forest growth using canopy metrics derived from airborne laser scanner data. \emph{Remote Sensing of Environment} \textbf{96}453--465.

\item Poudel K, Cao QV. 2013. Evaluation of methods to predict Weibull parameters for characterizing diameter distributions. \emph{Forest Science} \textbf{59}:243--252.

\item Ren Q, Banerjee S. 2013. Hierarchical factor models for large spatially misaligned data: A low-rank predictive process approach. \emph{Biometrics} \textbf{69}:19--30. 

\item Robert CP, Casella G. 2004. \emph{Monte Carlo Statistical Methods}. New York: Springer, Second Edition.

\item Roberts GO, Rosenthal JS. 2009. Examples of Adaptive MCMC. \emph{Journal of Computational and Graphical Statistics} \textbf{18}:349--367.

\item Robinson AP. 2004. Preserving correlation while modelling diameter distributions. \emph{Canadian Journal of Forest Research} \textbf{34}:221--232.

\item Salas C, Ene L, Gregoire TG, N{\ae}sset E, Gobakken T. 2010. Modelling tree diameter from airborne laser scanning derived variables: A comparison of spatial statistical models. \emph{Remote Sensing of Environment} \textbf{114}:1277--1285.

\item Schreuder HT Swank WT. 1974. Coniferous stands characterized with the Weibull distribution. \emph{Canadian Journal of Forest Research} \textbf{4}:518--523.

\item Sendak, P.E.,  Brissette, J.C., Frank RM. 2003. Silviculture affects composition, growth, and yield in mixed northern conifers: 40-year results from the Penobscot Experimental Forest, \emph{Canadian Journal of Forest Research}. \textbf{33}, 2116--2128.

\item Shepard N, Pitt MK. 1997. Likelihood analysis of non-Gaussian measurement time series. \emph{Biometrika} \textbf{84}:653--667.

\item Spiegelhalter D, Best N, Carlin B, van der Linde A. 2002. Bayesian measures of model complexity and fit. \emph{Journal of the Royal Statistical Society, Series B} \textbf{64}:583--639.

\item Wackernagel H. 2010. \emph{Multivariate Geostatistics: An Introduction with Applications. Third Edition.} New York: Springer. 

\item Wang M, Rennolls K. 2005. Tree diameter distribution modelling: introducing the logit--logistic distribution. \emph{Canadian Journal of Forest Research} \textbf{35}:1305--1313.

\item Weiskittel, AR, Hann DW, Kershaw JA, Vanclay JK. 2011. Forest growth and yield modeling. John Wiley \& Sons, West Sussex, UK.

\item Zhang, L., Gove JH, Liu C, Leak WB. 2001. A finite mixture of two Weibull distributions for modeling the diameter distributions of rotated-sigmoid, uneven-aged stands. \emph{Canadian Journal of Forest Research} \textbf{31}:1654--1659.

\end{description}

\newpage
\section*{Tables and figures}

\begin{table}[!ht]
\centering
\begin{tabular}{cccc}
  \hline
 &\multirow{3}{*}{} Non-spatial & Spatial & Spatial \\
                 & with LiDAR & without LiDAR & with LiDAR\\
  \hline
  $p_D$ &100.29 &102.65 & 214.01\\
  $DIC$ &-2967.79 &-3473.91 &-5572.22 \\
   \hline
\end{tabular}
\caption{Candidate model Deviance Information Criterion (DIC) and the effective number of parameters $p_D$.} 
\label{dic}
\end{table}

\begin{figure}[!h]
  \centering
  \includegraphics[width=12cm]{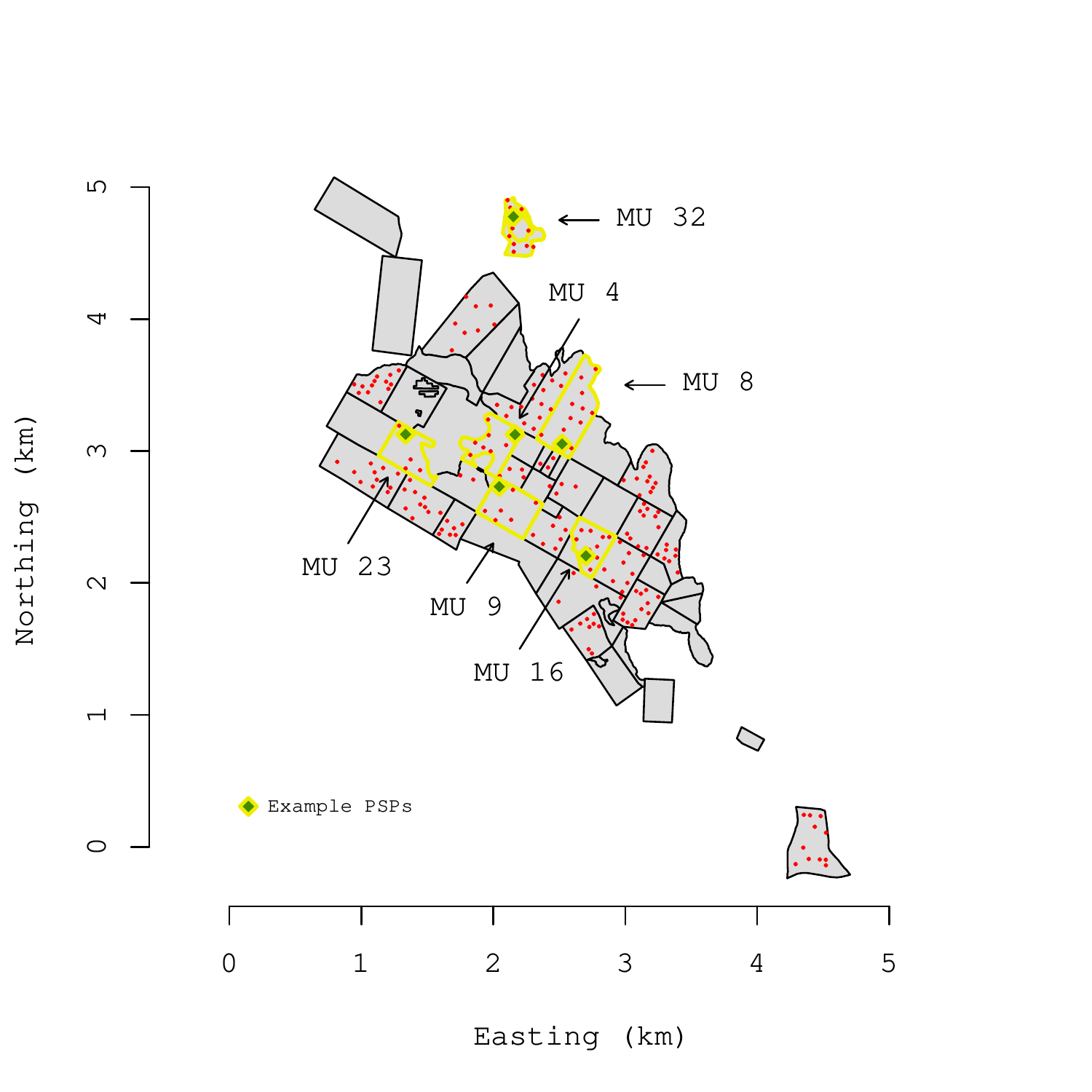}
  \caption{Map of PEF. PSPs highlighted in red. Example PSP referenced in Figure~\ref{fitted} colored in green. Black polygon boundaries outline different management units (MU). Select MUs have been labeled and highlighted in yellow.}
  \label{map}
\end{figure}

\begin{figure}[!ht]
 \begin{center}
   \subfigure[PSP 144 in MU 32]{\includegraphics[width=6cm]{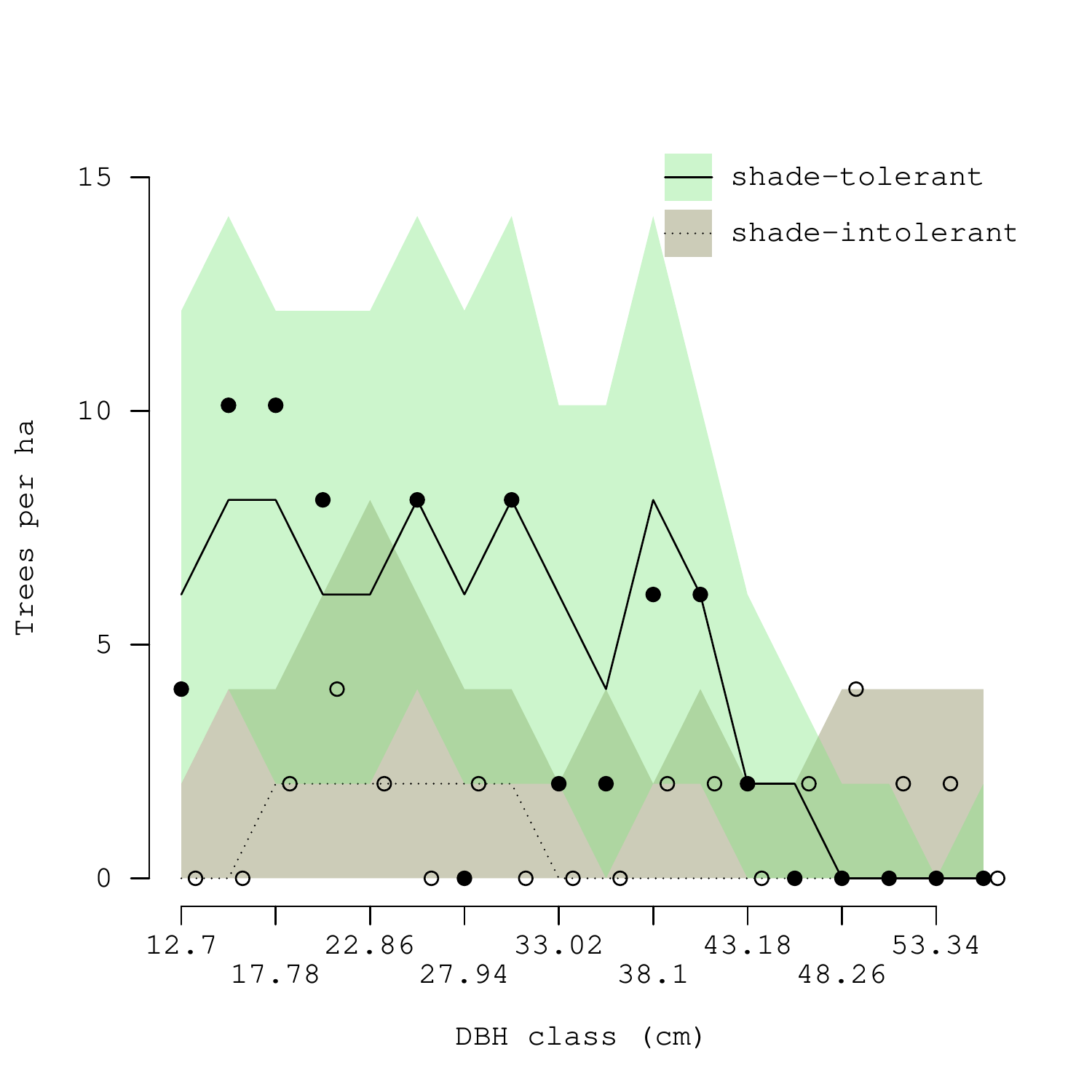}\label{h1}}
   \subfigure[PSP 192 in MU 8]{\includegraphics[width=6cm]{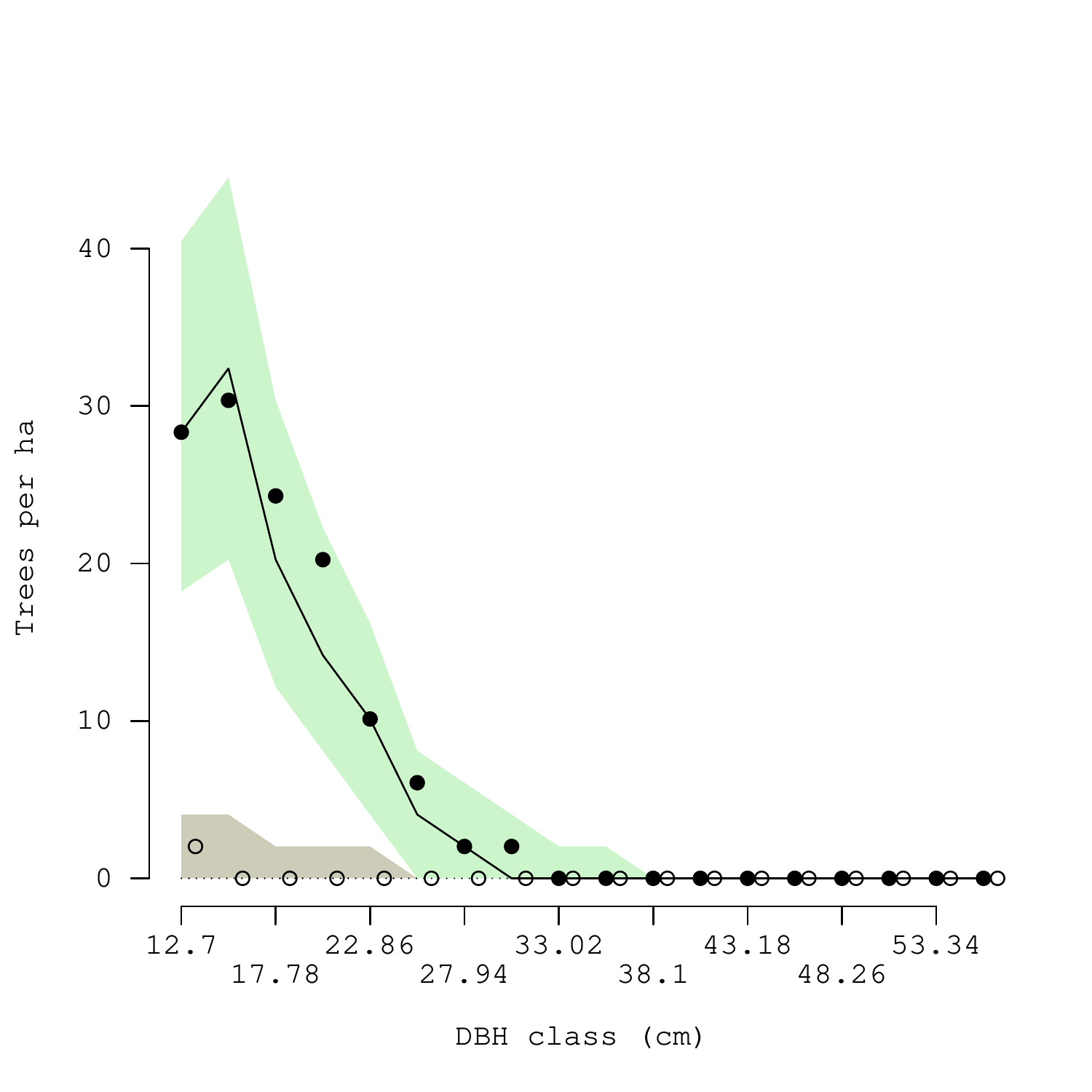}\label{h2}}
   \subfigure[PSP 41 in MU 16]{\includegraphics[width=6cm]{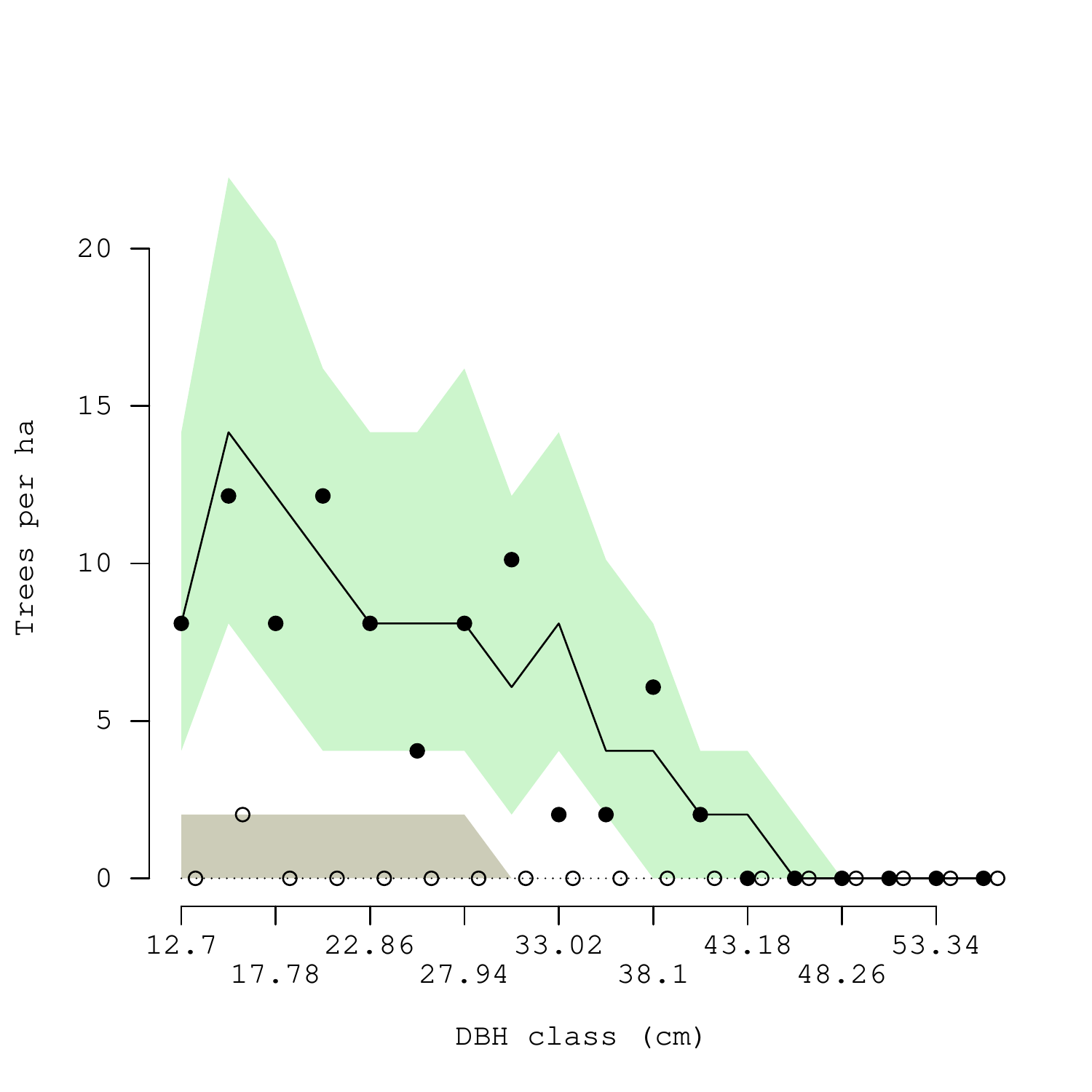}\label{h3}}
   \subfigure[PSP 208 in MU 9]{\includegraphics[width=6cm]{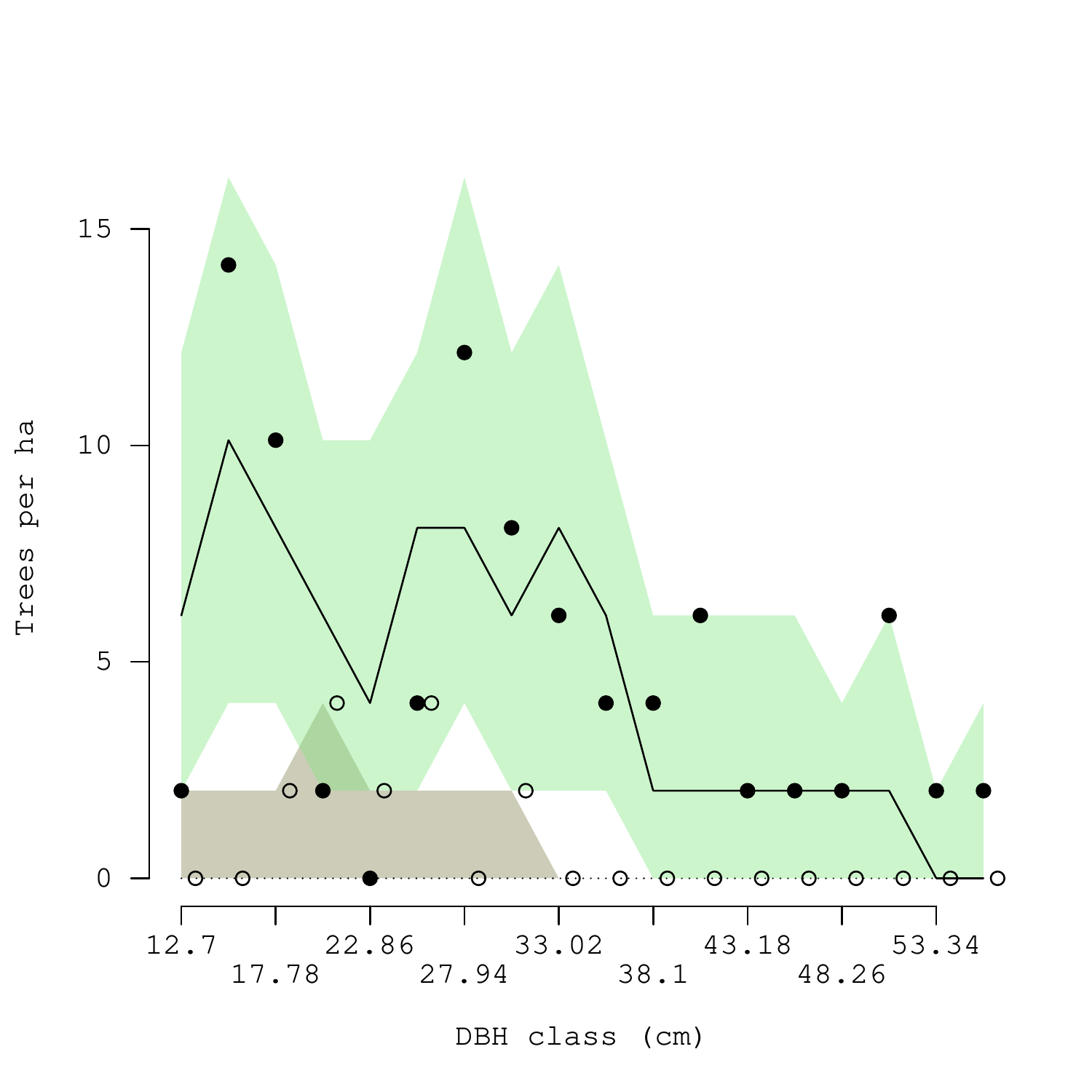}\label{h4}}
   \subfigure[PSP 151 in MU 4]{\includegraphics[width=6cm]{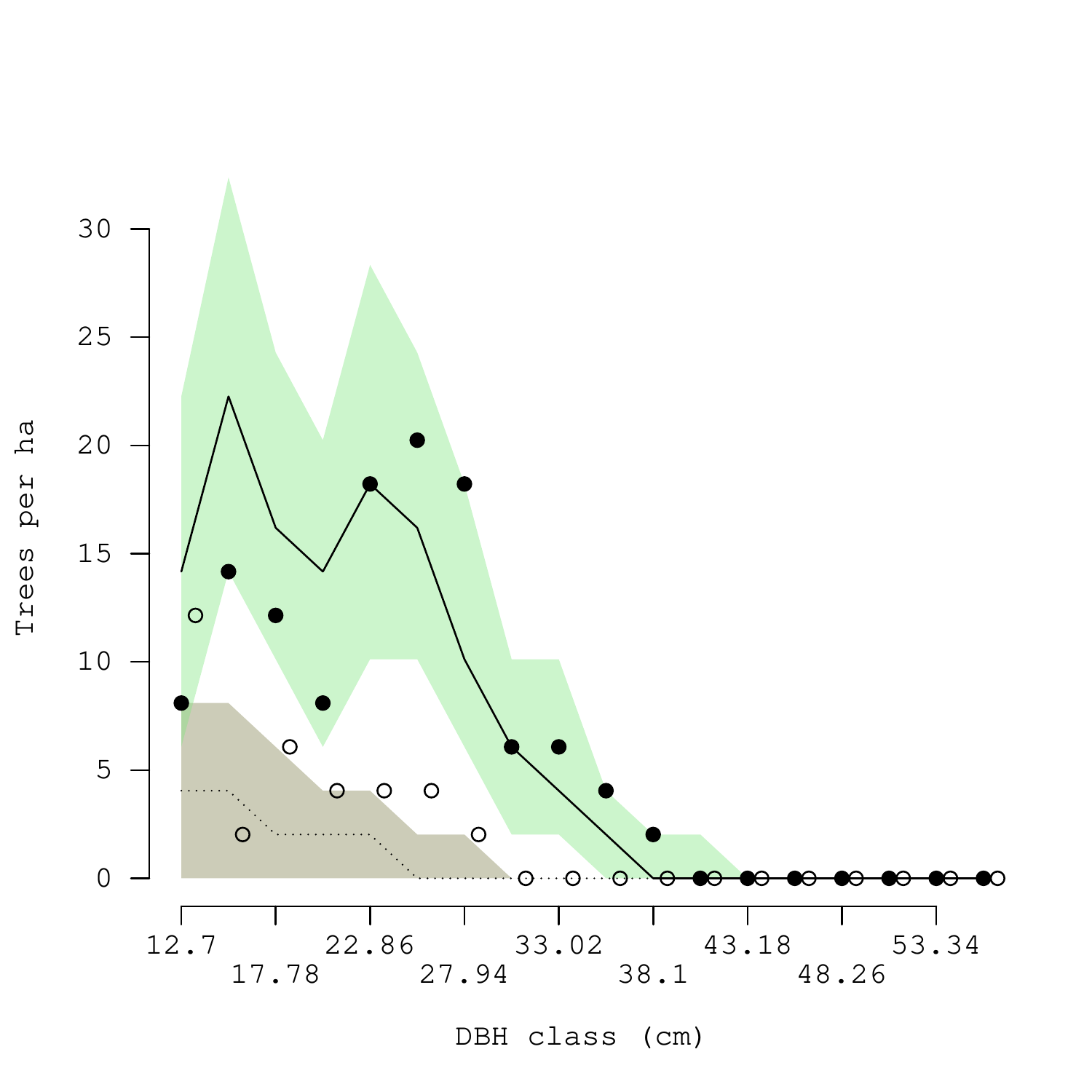}\label{h5}}
   \subfigure[PSP 80 in MU 23]{\includegraphics[width=6cm]{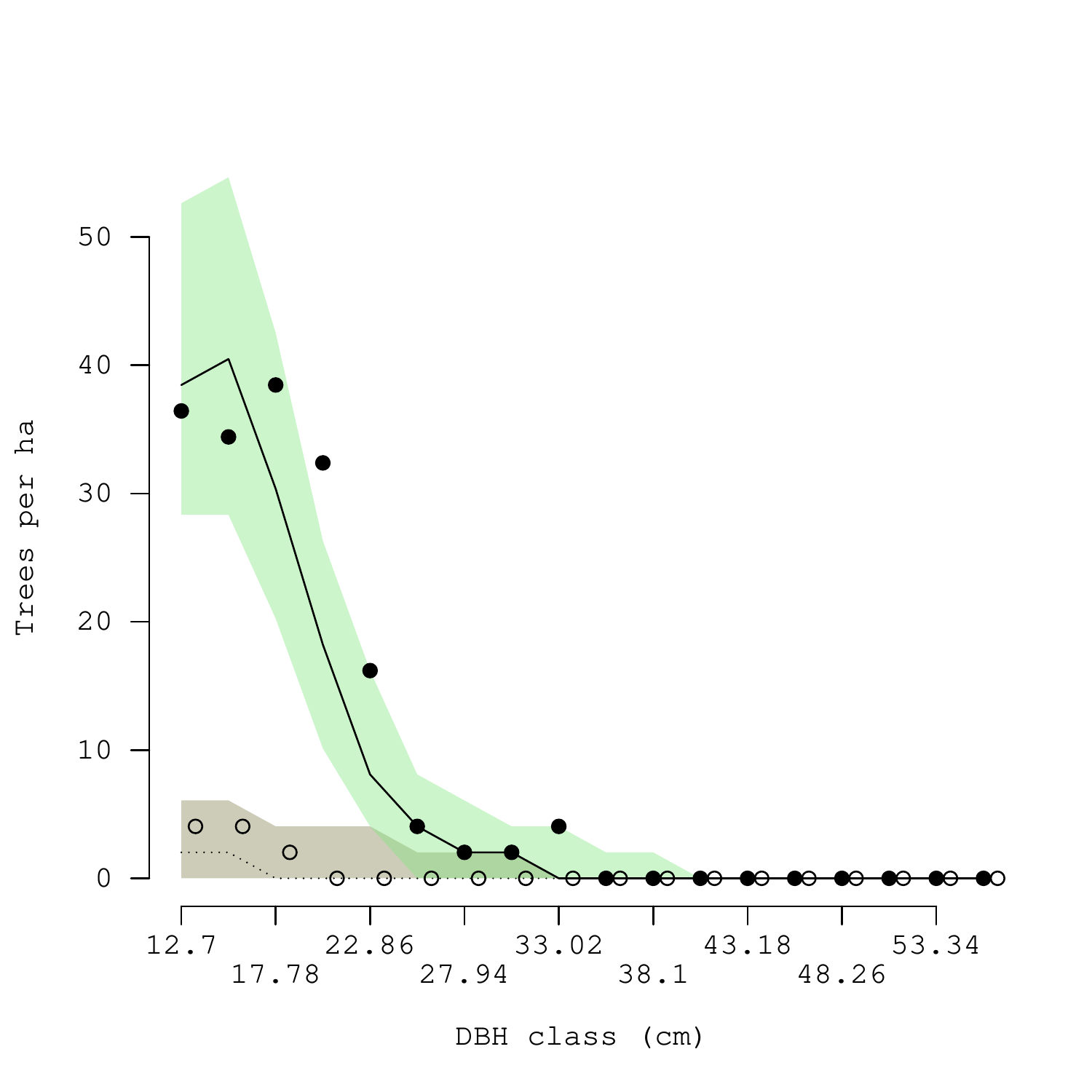}\label{h6}}
 \end{center}
 \caption{Observed and model~(\ref{Eq: Dynamic_Diameter_Class_Model}) fitted trees per hectare by diameter class for the six PSPs identified in Figure~\ref{map}. Solid black circles identify observed tree count for shade-tolerant and open circles correspond to observed tree count for shad-intolerant species. Solid lines with green envelops and dotted lines with gray envelops are the median and 95\% credible intervals for shade-tolerant and shade-intolerant respectively.}\label{fitted}
\end{figure}

\begin{figure}[!h]
  \centering
  \includegraphics[width=12cm]{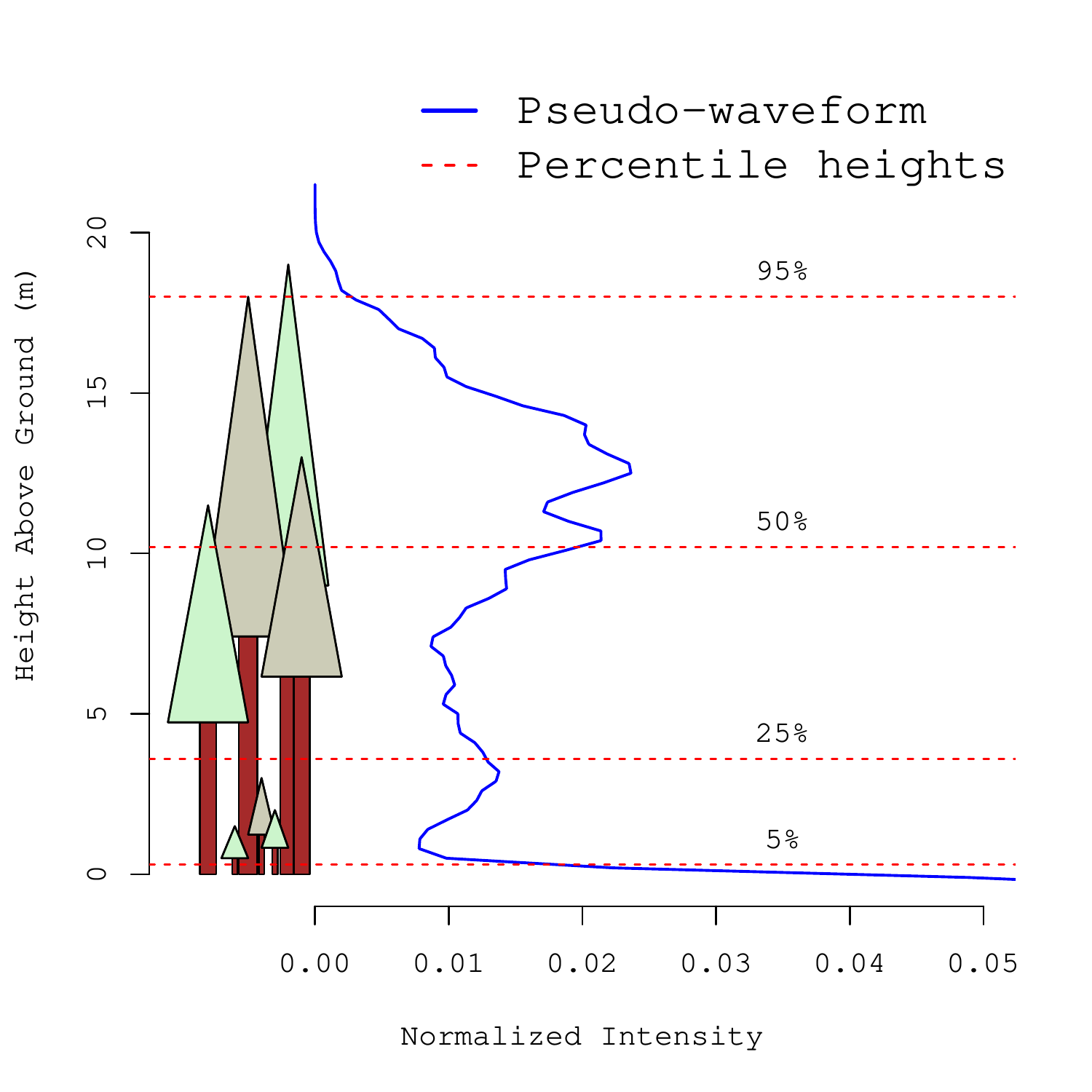}
  \caption{Illustration of a pseudo-waveform LiDAR signal, derived LiDAR percentile height metrics used as regressors, and vertical/horizontal forest structure at a generic location.}
  \label{lidar}
\end{figure}

\begin{figure}[!ht]
 \begin{center}
   \subfigure[LogS (shade-tolerant)]{\includegraphics[width=6cm]{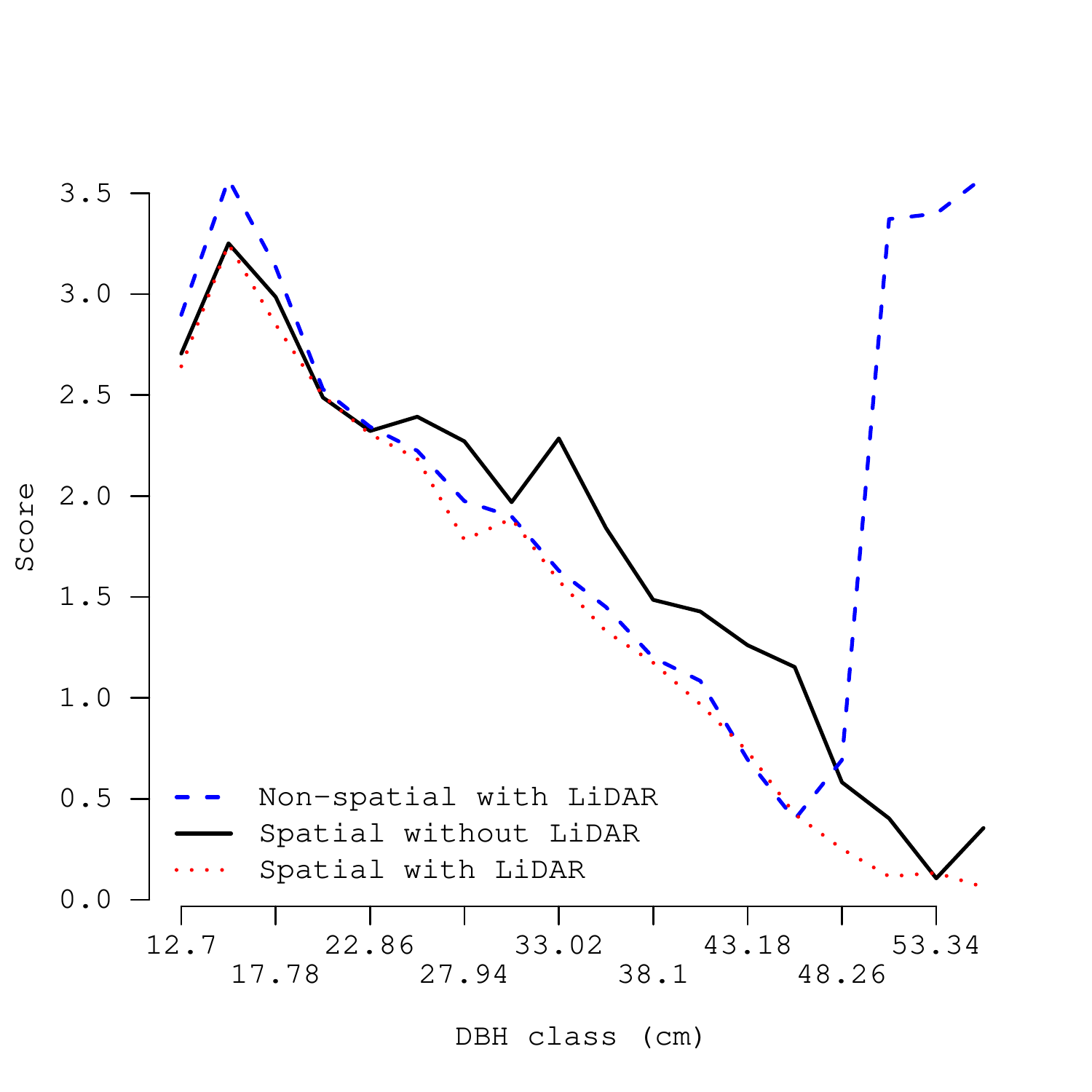}\label{logS1}}
   \subfigure[LogS (shade-intolerant)]{\includegraphics[width=6cm]{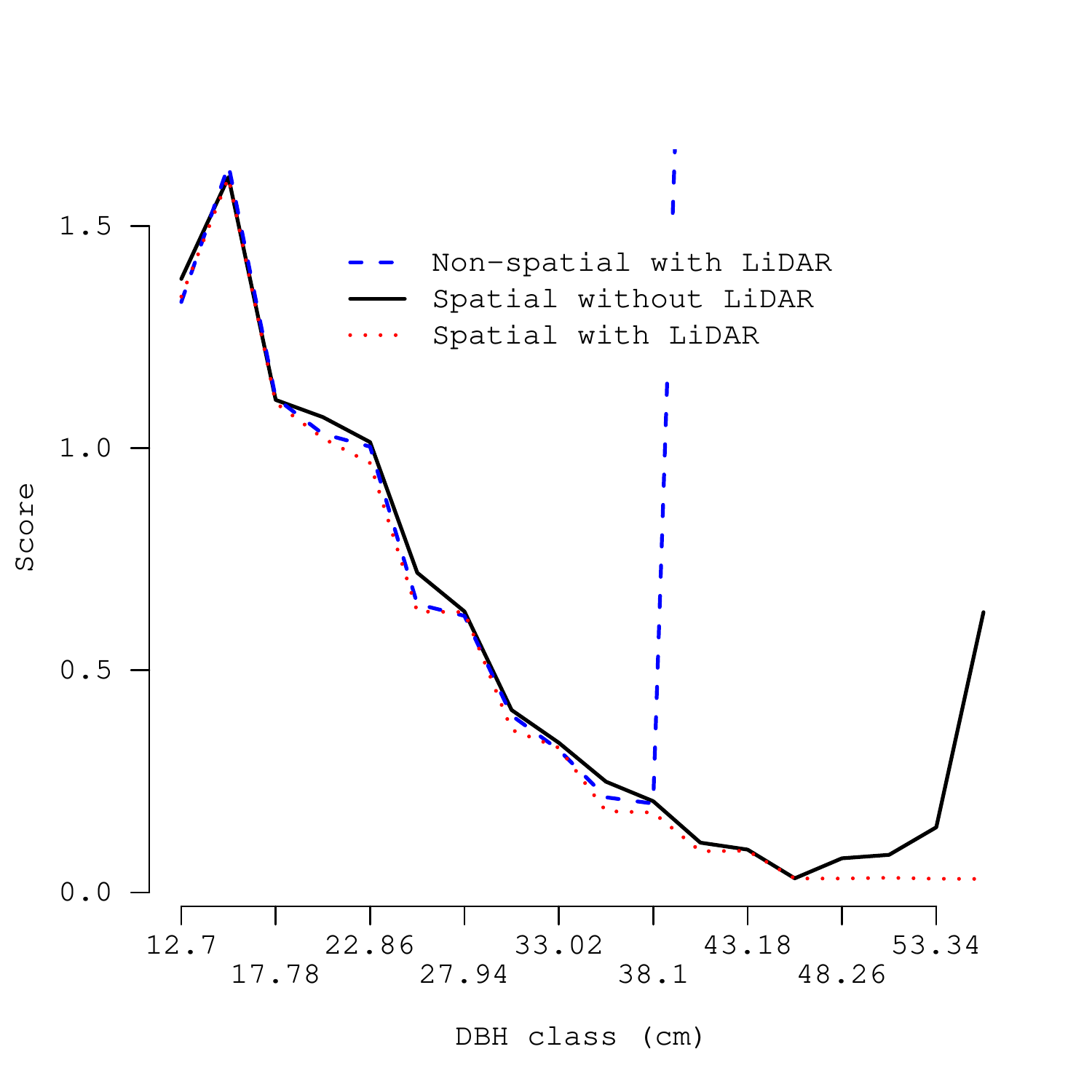}\label{logS2}}\\
   \subfigure[SES (shade-tolerant)]{\includegraphics[width=6cm]{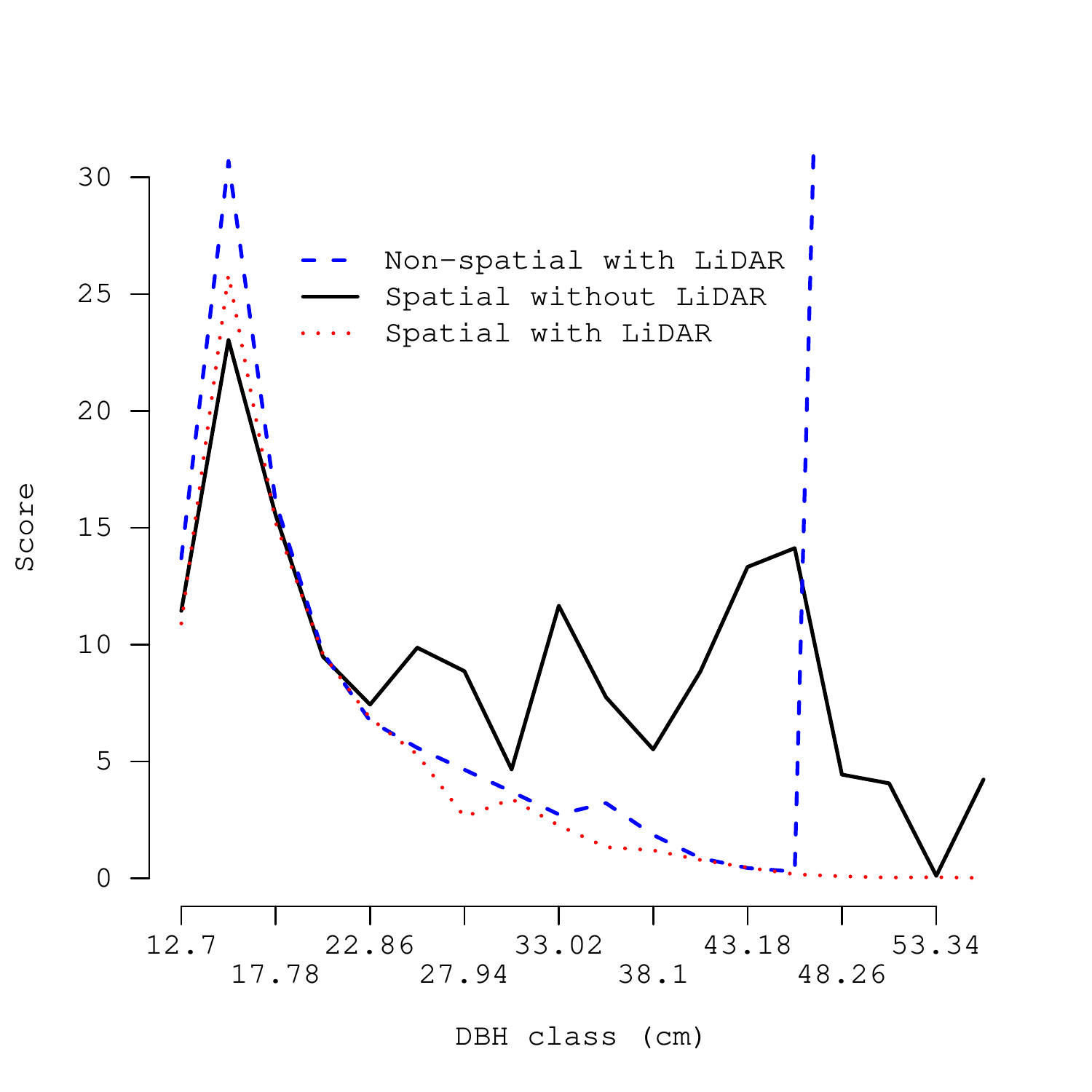}\label{SES1}}
   \subfigure[SES (shade-intolerant) ]{\includegraphics[width=6cm]{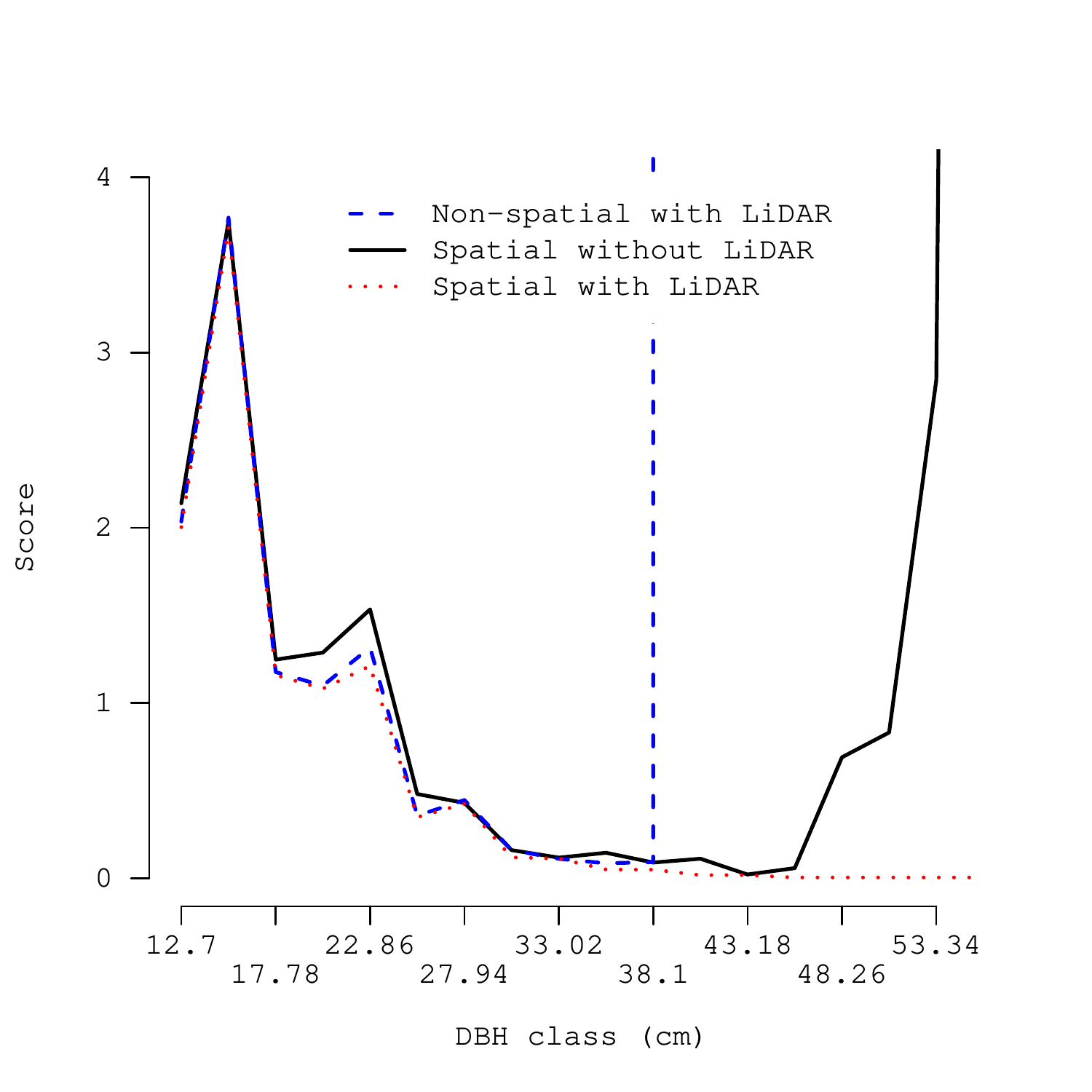}\label{SES2}}
   \subfigure[DSS (shade-tolerant)]{\includegraphics[width=6cm]{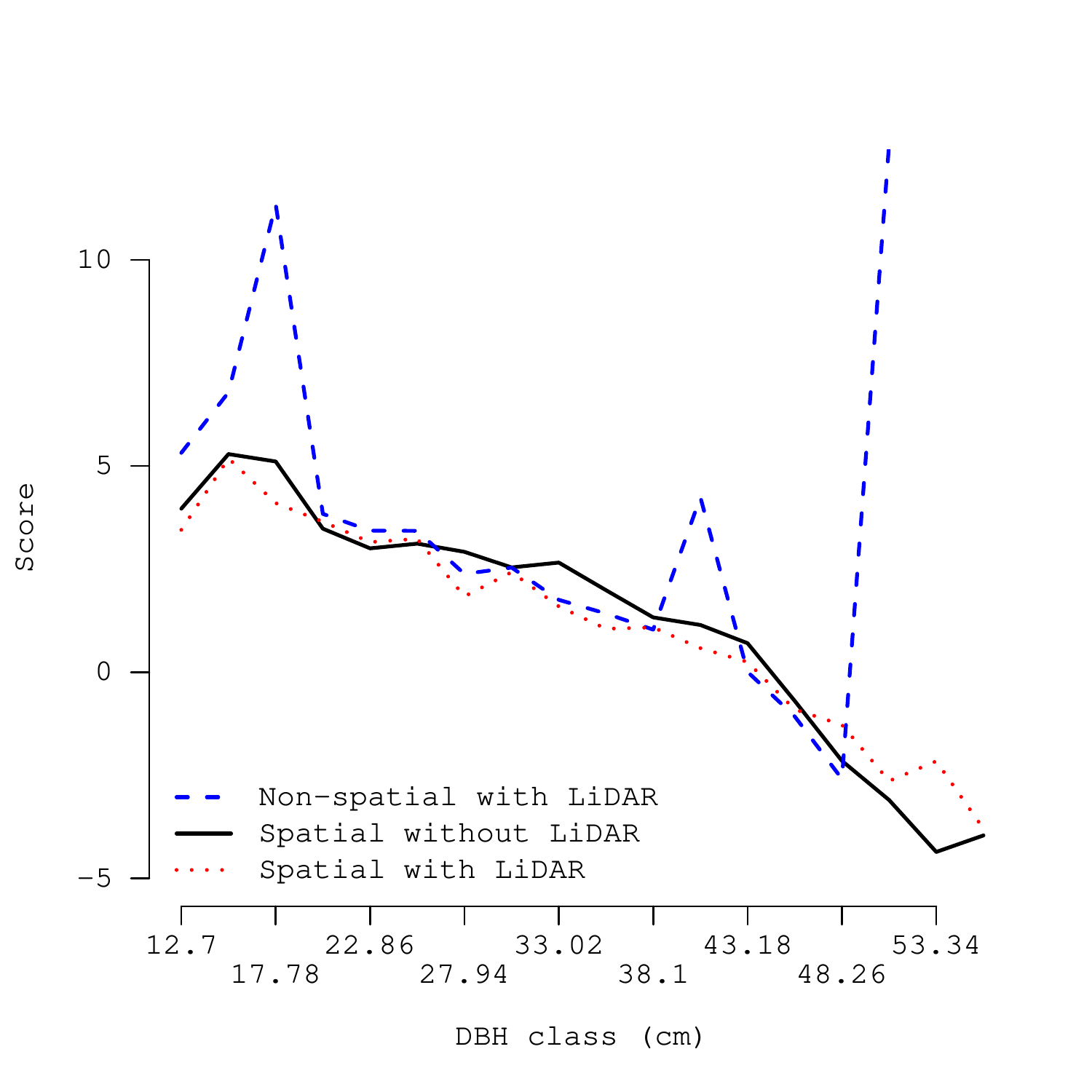}\label{DS1}}
   \subfigure[DSS (shade-intolerant) ]{\includegraphics[width=6cm]{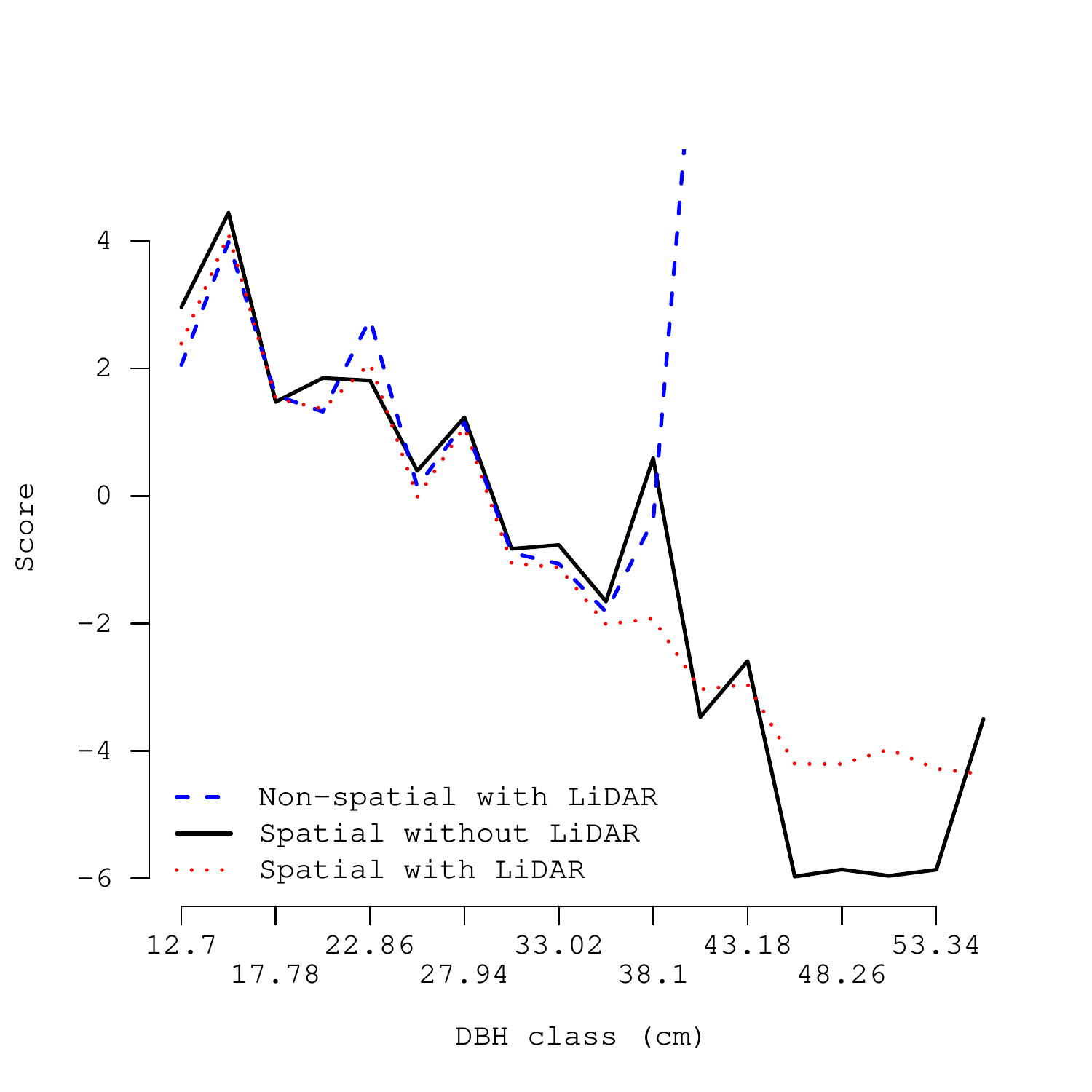}\label{DS2}}\\
 \end{center}
 \caption{Out of sample prediction performance using proper and strictly proper scoring rules for shade-tolerant and shade-intolerant.}\label{scores}
\end{figure}

\begin{figure}[!ht]
 \begin{center}
   \subfigure[$\beta_0$]{\includegraphics[width=6cm]{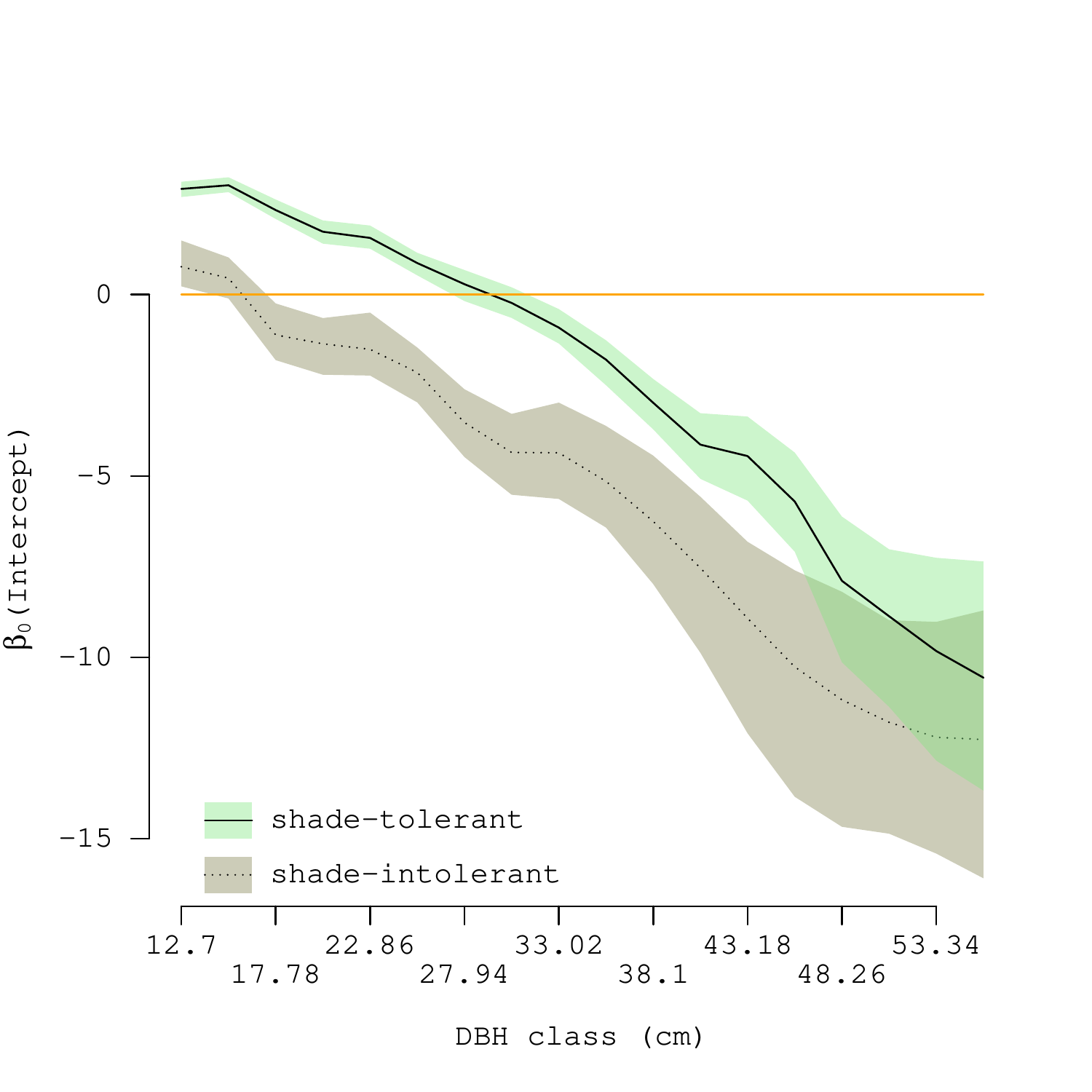}\label{b0}}\\
   \subfigure[$\beta_{P95}$]{\includegraphics[width=6cm]{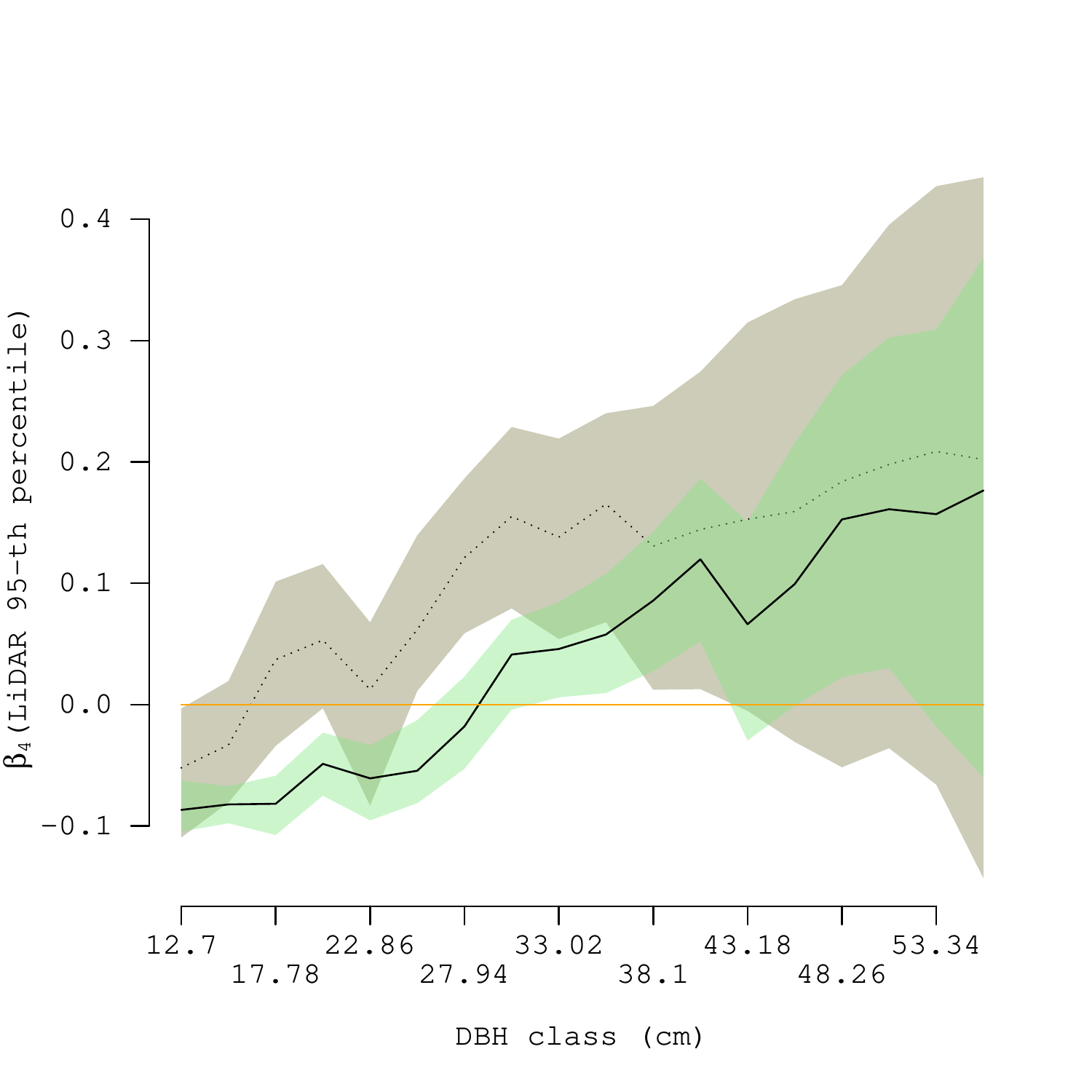}\label{b4}}
   \subfigure[$\beta_{P50}$]{\includegraphics[width=6cm]{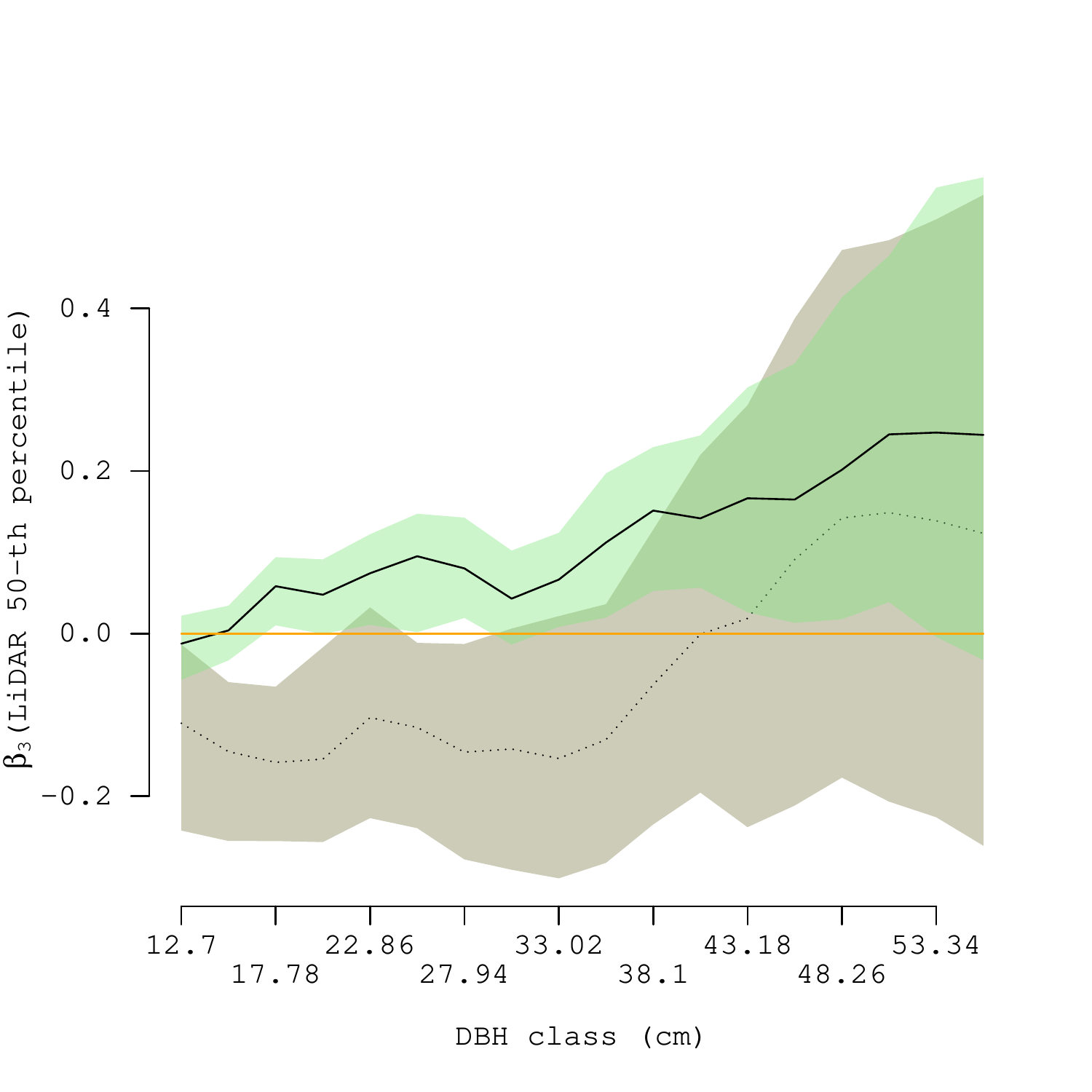}\label{b3}}\\
   \subfigure[$\beta_{P25}$]{\includegraphics[width=6cm]{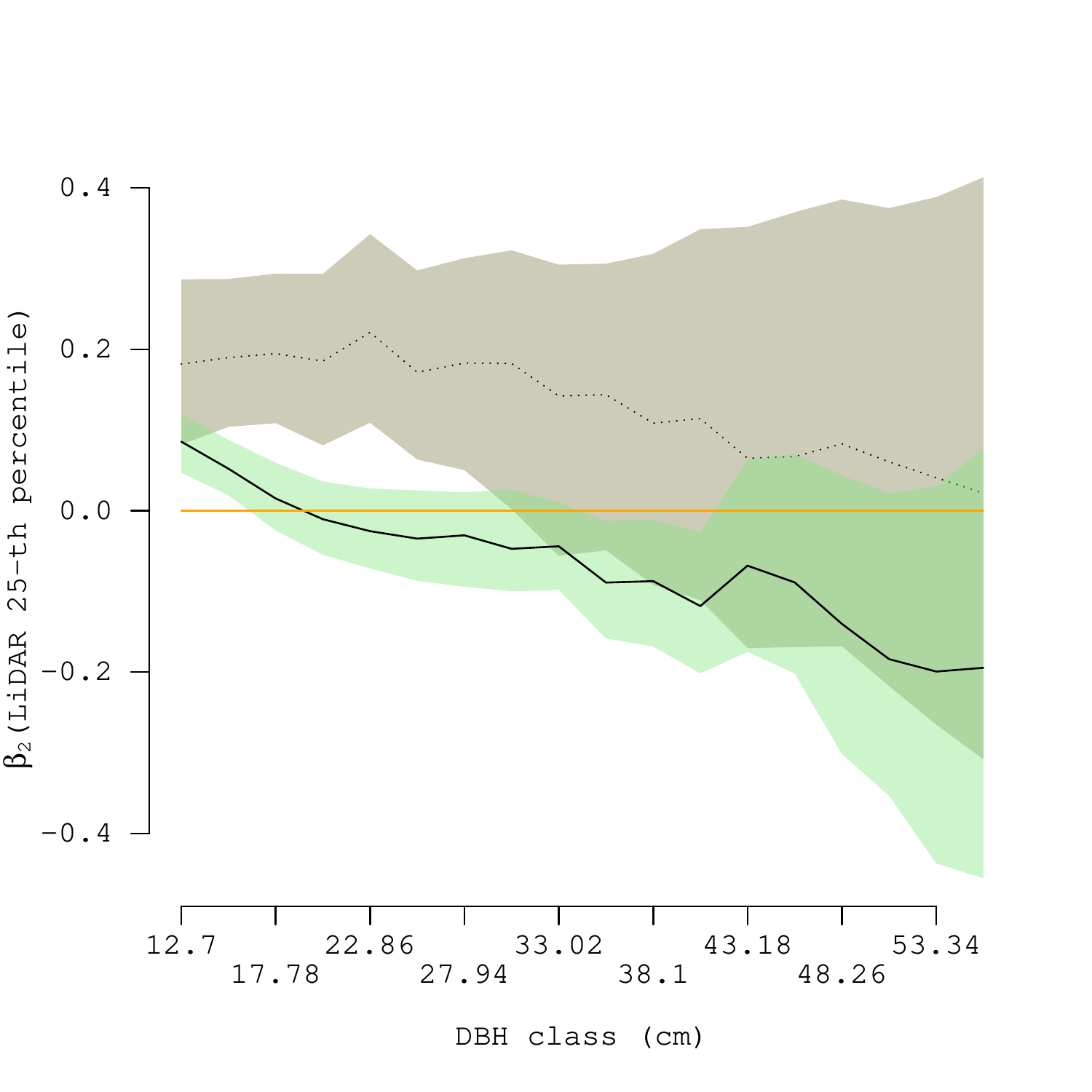}\label{b2}}
   \subfigure[$\beta_{P5}$]{\includegraphics[width=6cm]{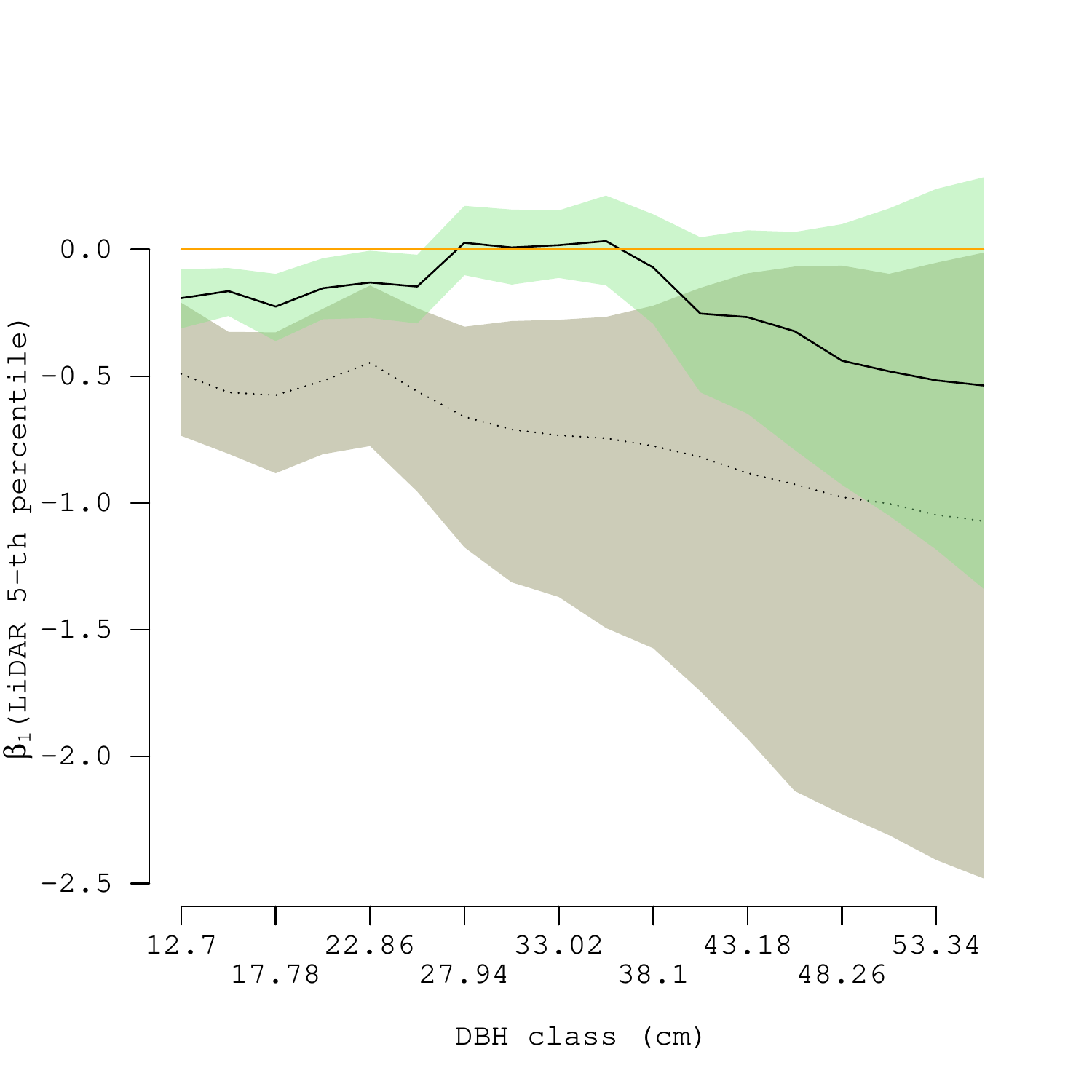}\label{b1}}
 \end{center}
 \caption{Model~(\ref{Eq: Dynamic_Diameter_Class_Model}) regression coefficients' posterior median and 95\% credible interval estimates for shade-tolerant and shade-intolerant.}\label{beta}
\end{figure}

\begin{figure}[!h]
  \centering
  \includegraphics[width=15cm]{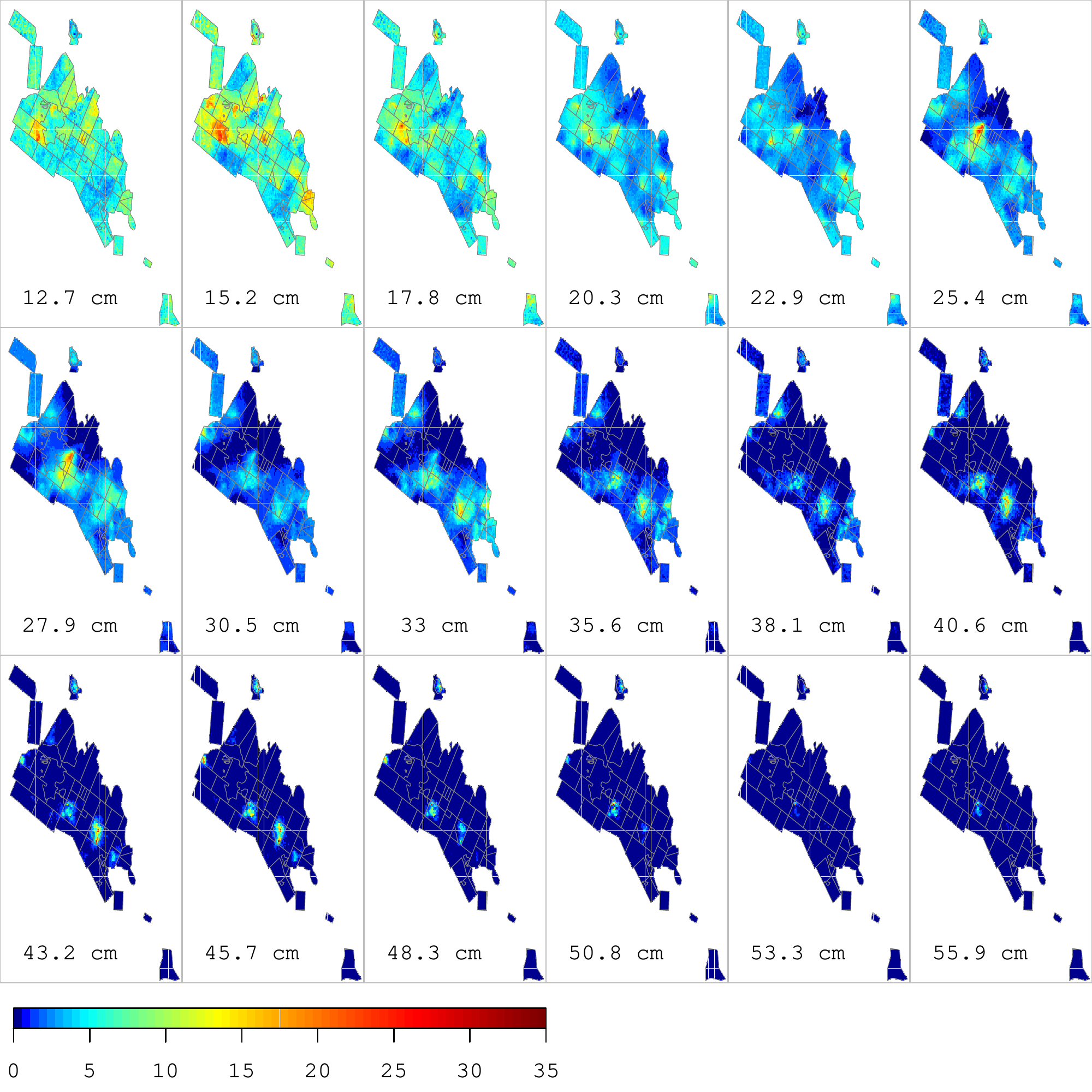}
  \caption{Model~(\ref{Eq: Dynamic_Diameter_Class_Model}) shade-tolerant trees per hectare by diameter class posterior predictive distribution medians.}
  \label{pred1}
\end{figure}

\begin{figure}[!h]
  \centering
  \includegraphics[width=15cm]{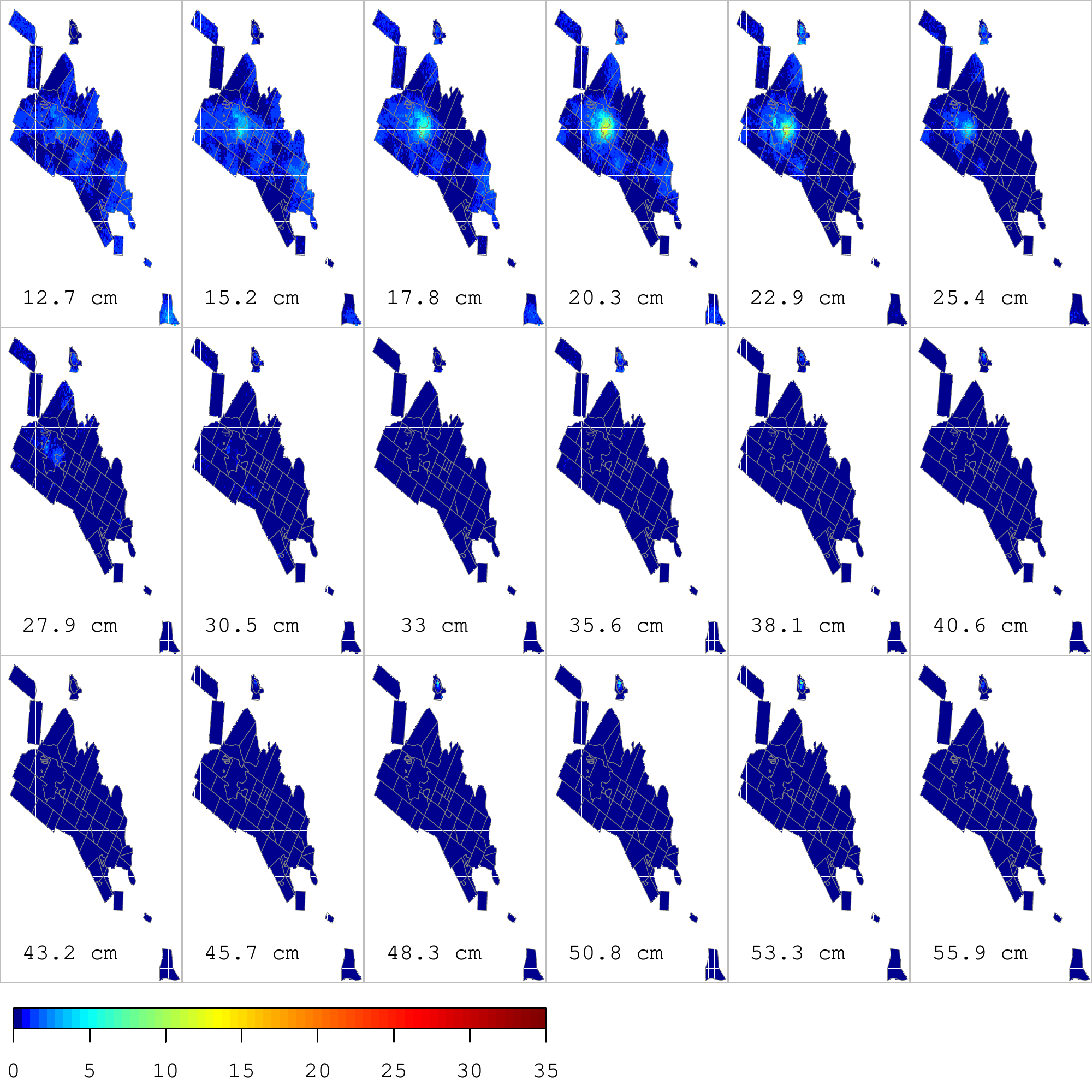}
  \caption{Model~(\ref{Eq: Dynamic_Diameter_Class_Model}) shade-intolerant trees per hectare by diameter class posterior predictive distribution medians.}
  \label{pred2}
\end{figure}

\begin{figure}[!h]
  \centering
  \includegraphics[width=15cm]{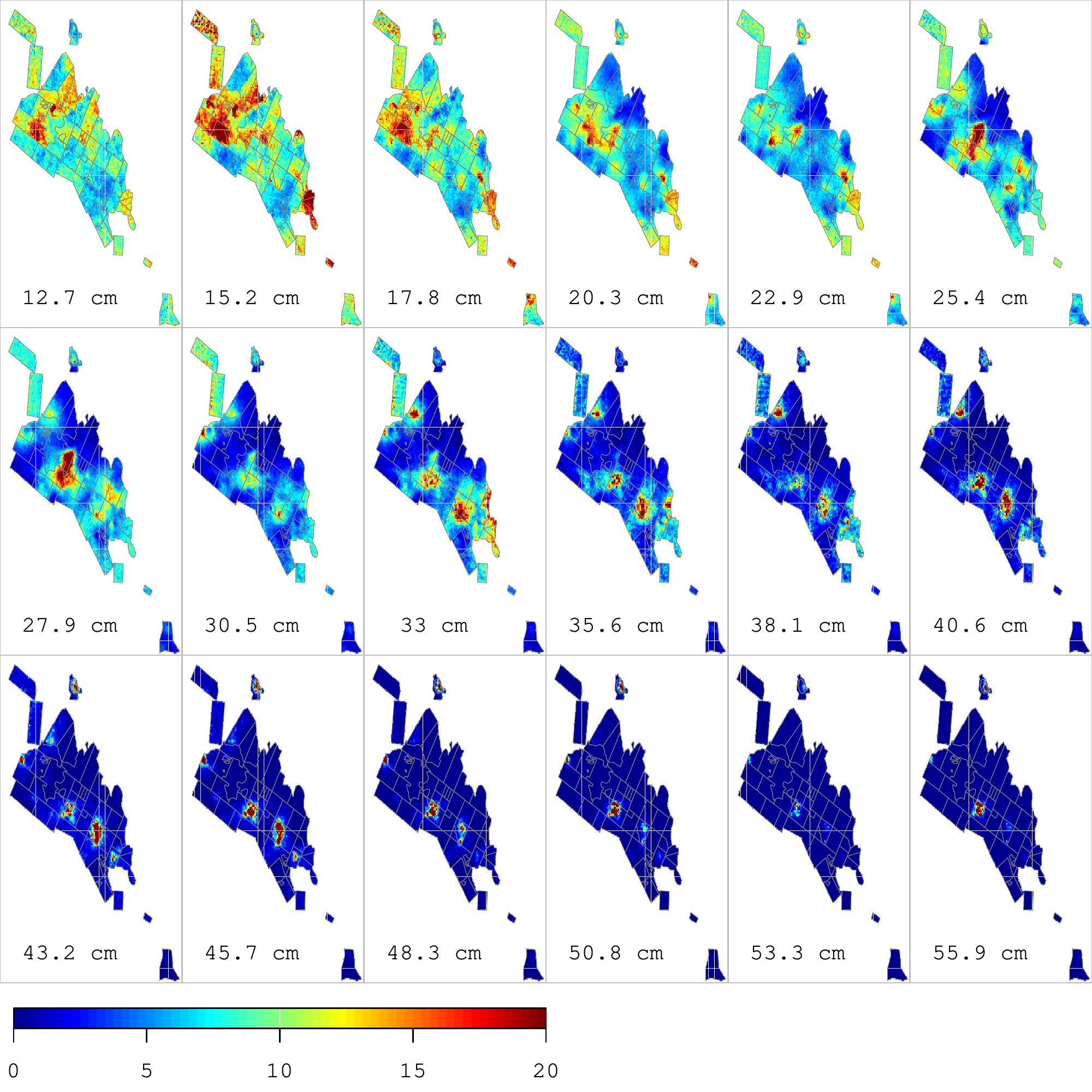}
  \caption{Model~(\ref{Eq: Dynamic_Diameter_Class_Model}) shade-tolerant trees per hectare by diameter class range between the lower and upper 95\% posterior predictive distribution credible intervals.}
  \label{pred1Rng}
\end{figure}

\begin{figure}[!h]
  \centering
  \includegraphics[width=15cm]{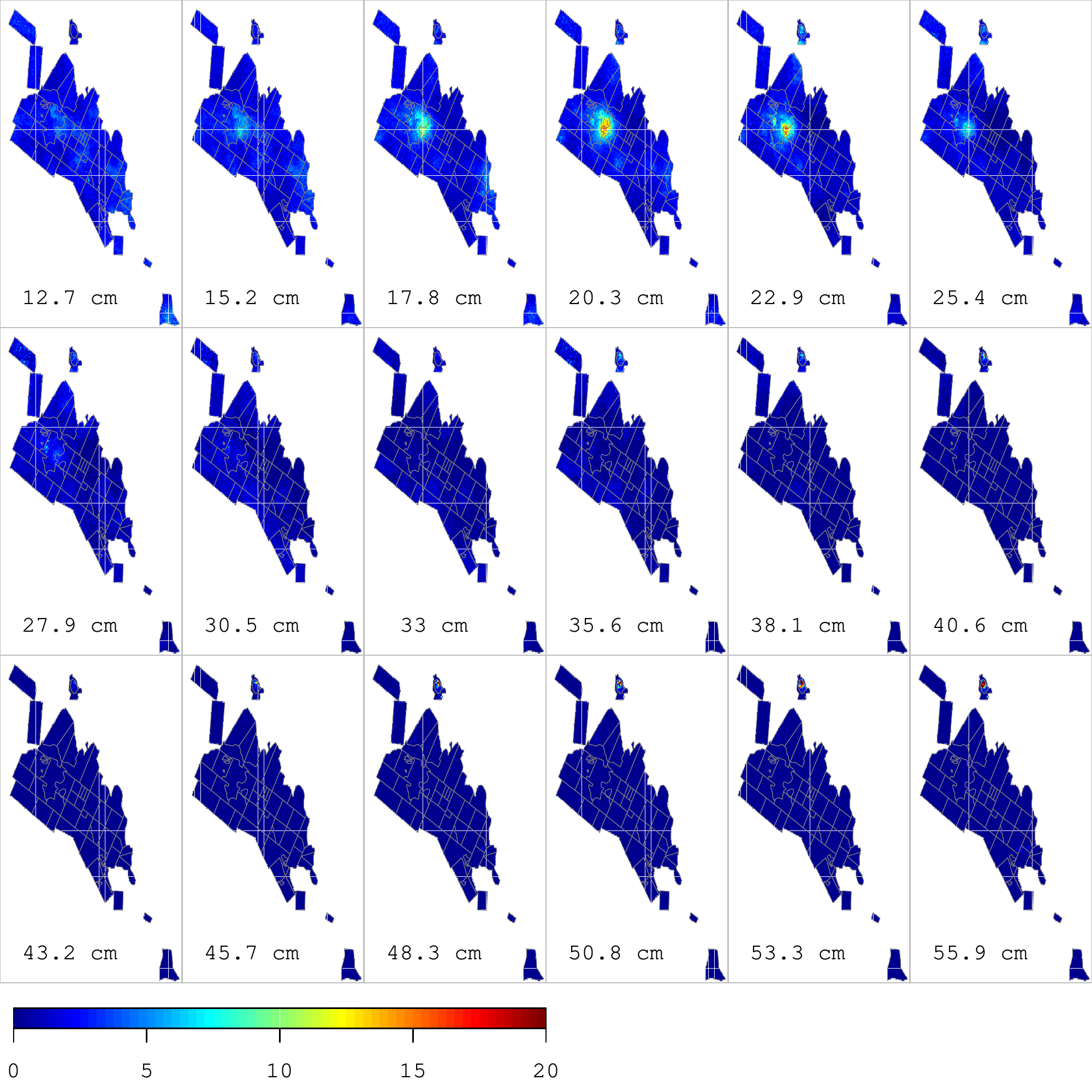}
  \caption{Model~(\ref{Eq: Dynamic_Diameter_Class_Model}) shade-intolerant trees per hectare by diameter class range between the lower and upper 95\% posterior predictive distribution credible intervals.}
  \label{pred2Rng}
\end{figure}

\end{document}